%% file: main.tex
\documentclass[12pt]{article}
\usepackage{fullpage}
\usepackage{mathrsfs}
\usepackage{float}
\usepackage{amsmath, amsfonts, amssymb, amsthm, amsbsy, amscd, bm, bbm}
\usepackage{url}
\usepackage{ifthen}

\usepackage{graphicx}
\usepackage{algorithm}
\usepackage{algorithmic}
\usepackage{hyperref}
\usepackage{balance}
\usepackage{subfig}
\usepackage{color}
\usepackage{paralist}
\usepackage{comment}
\usepackage{bm}
\usepackage{bbm}
\usepackage{cases}
\usepackage{balance}
\usepackage{kotex}
\usepackage{enumitem}
\usepackage[toc,page]{appendix}

\newtheorem{theorem}{Theorem}
\newtheorem{definition}{Definition}
\newtheorem{lemma}{Lemma}

\newtheorem{proposition}{Proposition}

\newcommand{\note}[1]{\textcolor{black}{#1}}

\newcommand{\perprice}{personalized pricing}
\newcommand{\nonprice}{common pricing}
\newcommand{\perpriceprob}{OPP}

\newcommand{\OPP}{\perpriceprob}

\newcommand{\linear}{p_c}
\newcommand{\linpol}{$\linear$}

\newcommand{\EXP}{\mathbb{E}}
\newcommand{\ep}{\hfill $\Box$}

\input{mymath}

\allowdisplaybreaks

\renewcommand{\paragraph}[1]{\smallskip \noindent{\bf #1.} }

\newcommand\defeq{\mathrel{:=}}

\title{Power of Bonus in Pricing for Crowdsourcing}
\author{
Suho Shin\thanks{LINE Plus Corporation}
\and
Hoyong Choi\thanks{Department of Electrical Engineering, Korea Advanced Institute of Science and Technology}
\and
Yung Yi\footnotemark[2]
\and
Jungseul Ok\thanks{Department of Computer Science and Engineering and Graduate School of Artificial Intelligence, Pohang University of Science and Technology}
}

\begin{document}

\maketitle
\begin{abstract}
We consider a simple form of pricing for a crowdsourcing system, where pricing policy is published a priori, and workers then decide their task acceptance. Such a pricing form is widely adopted in practice for its simplicity, e.g., Amazon Mechanical Turk, although additional sophistication to pricing rule can enhance budget efficiency. With the goal of designing efficient and simple pricing rules, we study the impact of the following two design features in pricing policies: (i) personalization tailoring policy worker-by-worker and (ii) bonus payment to qualified task completion. In the Bayesian setting, where the only prior distribution of workers' profiles is available, we first study the Price of Agnosticism (PoA) that quantifies the utility gap between personalized and common pricing policies. We show that PoA is bounded within a constant factor under some mild conditions, and the impact of bonus is essential in common pricing. These analytic results imply that complex personalized pricing can be replaced by simple common pricing once it is equipped with a proper bonus payment. To provide insights on efficient common pricing, we then study the efficient mechanisms of bonus payment for several profile distribution regimes which may exist in practice. We provide primitive experiments on Amazon Mechanical Turk, which support our analytical findings.
\end{abstract}

\maketitle

\input{intro}
\input{related}

\input{model}

\input{result}

\input{simulation}

\input{experiment}
\input{conclusion}
\bibliography{ref}
\clearpage
\appendix
\appendixpage
\input{appendix}

\clearpage

\end{document}

%% file: mymath.tex
\newcommand{\set}[1]{\ensuremath{\mathcal #1}}
\renewcommand{\vec}[1]{\bm{#1}}

\newcommand{\separator}{
  \begin{center}
    \rule{\columnwidth}{0.3mm}
  \end{center}
}

\def\Bl{\Bigl}
\def\Br{\Bigr}
\def\lf{\left}
\def\ri{\right}

\newcommand{\bprob}[1]{\mathbb{P}\Bl[ #1 \Br]}
\newcommand{\prob}[1]{\mathbb{P}[ #1 ]}
\newcommand{\expect}[1]{\mathbb{E}[ #1 ]}
\newcommand{\bexpect}[1]{\mathbb{E}\Bl[ #1 \Br]}

\newcommand{\beq}{\begin{eqnarray*}}
\newcommand{\eeq}{\end{eqnarray*}}
\newcommand{\beqn}{\begin{eqnarray}}
\newcommand{\eeqn}{\end{eqnarray}}
\newcommand{\bemn}{\begin{multiline}}
\newcommand{\eemn}{\end{multiline}}

\def\ep{\hfill $\Box$}

%% file: intro.tex
\section{Introduction}
\label{sec:intro}

Crowdsourcing system is a popular tool to solve problems 
which involve huge amount of simple tasks,
where the tasks are electronically distributed to numerous workers who are willing to perform the tasks at low cost. 
Once a task is given to workers, 
error is often common even among those who are willing due to low payment, tedium in tasks, and/or abundant spammers.
To handle this, on one hand, 
\note{a number of post-processing methods are developed} to denoise dataset
with some statistical inference techniques such as expectation maximization or belief propagation are applied to infer the correct answer to the task \cite{Inference:1,Inference:2,Inference:4}.
However, there should be unavoidable limitations in guaranteeing 
a target precision when worker inputs are highly erroneous \cite{FalseReport:1,FalseReport:2}.
On the other hand, there have been an extensive line of studies on {\em worker quality control},
where workers are incentivized 
to submit better answers by the corresponding higher payment
\cite{Incentivehelp:1,Incentivehelp:2}.  
Many experiments reveal that both the quantity and the quality of participants' labels improve  
under the smart incentivization \cite{QuantityImprove:1,QuantityImprove:2}.


In many microtask crowdsourcing systems, a simple and eidetic pricing mechanism without bidding process \note{is} widely adopted due to the volatility of the incoming workers and characteristic of the tasks to be solved.
For example in Amazon Mechanical Turk (MTurk), a task requester publishes a pricing rule a priori, based on which each worker decides on the task acceptance. Such workers are finally admitted to the system, if the requester has budget to pay them for their work.
The following ideas are natural as possible ways to increase the requester's utility:  
(i) \note{{\em personalized pricing}} that offers different rule to determine 
price for each individual worker and 
(ii) giving additional \note{{\em bonus}}\footnote{\note{In here, we indicate bonus payment to be an ex-post reward that the workers are paid based on their quality of contributions. Note that this is distinguished from base payment which is given to the workers immediately once he accepts to participate in the task, where we present the concrete definitions in Secion~\ref{sec:model}.}} to workers with more qualified task completion.
\note{{\em Personalized pricing}} would be  
superior to the
pricing without personalization (which we call {\em common pricing} throughout this paper).
However, {\perprice} is obviously more complex and even cannot be adopted in some systems such as Mturk. A simple option of giving \note{{\em bonus}} to qualified workers in both {\perprice} and {\nonprice} should provide more power in controlling workers' behaviors which helps in increasing the requester's utility.
However, it has been underexplored how much gain personalization and/or bonus payment in pricing policies actually provide to the task requester. 


The key message which we claim in this paper 
is that common pricing is enough when it is equipped with an appropriate choice of bonus payment, so as to catch two rabbits of simplicity and efficiency simultaneously.
The main contributions to support our claim are summarized as follows:

\vspace{0.1cm}
\begin{compactenum}[(a)]

\item {\bf Price of Agnosticism.} We first study {\em Price of Agnosticism (PoA)} of a common pricing, which quantifies the gap between the optimal personalized pricing and a given common pricing (see Theorem~\ref{thm:ex_post}). This PoA is expressed as $(\delta,\gamma)$-approximation, where $\delta \leq 1$ and $\gamma \geq 1$ correspond to the degree of approximations in terms of achieved utility and used budget, respectively, i.e., achieving $\delta U^\star_p(B)$ with $\gamma B$ budget, where  $U^\star_p(B)$ is the maximum utility by the optimal personalized pricing for a given budget $B.$
We prove that there exists a 
common pricing which obtains the PoA $(1-o(1),O(1))$ for a large number of workers under some canonical scenarios.

\item {\bf Power of Bonus.} 
We next study {\em Power of Bonus (PoB)} that explains the role of bonus payment in common pricing as a simple, yet powerful mechanism (see Theorem~\ref{thm:pad_bonus}). 
We prove the necessity of bonus in common pricing in the sense that, without bonus, there always exists a worker profile distribution and arrival order under which any common pricing without bonus payment cannot achieve $(O(1), o(n))$-approximate to optimal pricing policy.

\item {\bf Optimal Common Pricing.}
For some canonical profile distribution regimes, we find the optimal structure of common pricing policies.
Since the profile distribution can be interpreted as a consequence developed by the characteristic of task the requester is trying to solve along with the corresponding behavior of the workers, these results give some insights in designing the pricing policies with respect to the requester's task.

\item {\bf Experiments.}
To validate our findings and draw practical implications,
we numerically analyze the various profile distribution regimes and efficient structure of pricing policy, and execute a real-world experiment on Amazon Mechanical Turk, a popular crowdsourcing platform. The results from experiments verify the theoretical result on PoA, PoB, and the efficiency of some bonus structures, and it also gives some intuitions on how to build an efficient pricing policy given the requester's utility function.


\end{compactenum}
\vspace{0.1cm}

To sum up, our analytical and experimental results imply that a simple common pricing equipped with bonus payment is enough with no need of personalized pricing, and provide some useful insights in efficient form of pricing policies with respect to the crowdsourcing scenario.

In Section~\ref{sec:model}, we introduce our model and pricing mechanism, and in Section~\ref{sec:poa}, we provide our main analytical results on PoA, PoB along with some case studies on worker regime. In Section~\ref{sec:simulation}, we provide the numerical results to verify the theoretical findings, and in Section~\ref{sec:experiment}, we provide the experimental results based on the real-world crowdsourcing data on Mturk. All the proofs are provided in Appendix.

%% file: related.tex
\subsection{Related Work}
\label{sec:related}

In this section, we provide an overview of literatures related to 
pricing mechanisms for microtask crowdsourcing systems, in which 
we want to incentives quality workers under budget constraints.

\paragraph{Pricing mechanisms with budget feasibility}
We first present the work on pricing mechanism in crowdsourcing that is based on a procurement auction under budget feasibility constraint, i.e. budget feasible mechanism design. 
\cite{BudgetFeasible:Singer10} initiatively shows that for a general buyer utility function, standard mechanism design ideas such as VCG mechanism along with its variants cannot guarantee a constant-approximation to an optimal mechanism, while it becomes possible for certain classes of submodular utility functions. \cite{BudgetFeasible:Anari13} study the analogous problem under large market assumption, and \cite{BudgetFeasible:Bei10} study the same problem under subadditive function in both prior-free and Bayesian framework.

Despite of the advantages of sealed-bid mechanism, 
they are rarely adopted in microtask crowdsourcing platforms 
due to the complex bidding process. However, 
posted pricing is widely used due to its simplicity \cite{Posted:Amanatidis19, Posted:Sun14, Posted:Sekar16, Posted:Hu17}. In this context, \cite{Posted:Balkanski16} define a Bayesian budget-feasible posted pricing, where given a prior knowledge on individual worker's cost distribution, the buyer tries to maximize her expected utility within a fixed ex-post budget constraint.  \cite{qualityaware:Han16, qualityaware:Han18} study the Bayesian posted pricing mechanism under the prior knowledge in joint distribution of worker's cost and quality, and formulate the problem as a budget minimization problem under robust quality constraint. 
We study a variant of budget feasible posted price mechanism, where the requester incentivizes the participants by giving a bonus payment based on the observed quality in addition to a participation fee.

\paragraph{Quality-based pricing and strategic behavior of workers}
Since we integrate the concept of quality-based payment in posted price mechanism, we introduce some works studying on the strategic behavior of workers against the performance-based contract in labor market along with its difference from our model, and discuss why our model practically fits in our application.
The literature of principal-agent problem~\cite{contract:laffont,contract:babaioff, Contract:Carroll15, contract:lervy}
addresses the question of incentivizing strategic workers based on their observed output, where the main focus is to design a contract to elicit costly efforts from strategic agents. In here, each agent strategically decides an effort level to exert on the task to maximize own payoff, where higher effort induces large opportunity cost but possibly large reward, and lower effort would not make much reward but it takes a small opportunity cost. \note{There exists a number of works studying the crowdsourcing system in the context of principal-agent problem~\cite{contract:ho, contractcs:bhat, contractcs:chen, contractcs:ma, QualityImprove:Ho15}.}
Meanwhile, there have been reported some empirical evidences that the workers in microtask crowdsourcing platforms do not strategize their effort level or induced quality, but only decide whether or not to participate in the requested task~\cite{evidence:jeon, evidence:kraut, evidence:mason}.
In this context, the authors in \cite{behavioral:easley,behavioral:ghosh} capture these characteristics of worker behavior, namely endogenous participation and exogenous quality, to model the real-world crowdsourcing platform and reveal the structure of efficient mechanism. Our work also considers an analogous intrinsic worker behavior, but in a budget-constrained scenario under posted price mechanism to capture the characteristic of the microtask crowdsourcing platform.

\paragraph{Price discrimination}
Our main message is that differentiating pricing policy between workers is not necessary if its equipped with proper bonus payment in microtask crowdsourcing platform.
In this sense, we explain the concept of price discrimination, why it is needed in terms of system utility, and introduce some works on studying the performance of system under less price discrimination.
Firstly, price discrimination refers to the selling strategy that changes its offering price for different individual customers based on their observable features such as willingness to pay, or the utility they provide to the requester. It is an effective tool, or sometimes inevitable for recruiting qualified participants while filtering out poor ones which leads to higher requester utility, as used in extensive mechanism design literature~\cite{Posted:Chawla09, Posted:Balkanski16, qualityaware:Han16, OnlinePosted:Han17, discrimination:cummings}.
However, it is also widely known that differentiating pricing policy would possibly result in a negative effect on platform in long-term manner~\cite{Discrimination:Anderson10}. 
In this context, there exists a line of works on studying the performance of system without or less discrimination under monopoly provider scenario~\cite{discrimination:elmachtoub, discrimination:bergemann, discrimination:bergemann2}, or under the presence of network externalities~\cite{discrimination:huang}.
We also address how the requester can effectively allocate the budget to workers without differentiating the pricing policy in microtask crowdsourcing platform, and reveal that if there exists no discrimination at all, then the performance of requester drops significantly, however, with {\em implicit} price discrimination occurred by proper quality-based bonus payment, the requester can achieve a near-optimal performance compared to pricing policy with both {\em explicit} discrimination via differentiation and implicit discrimination.

\paragraph{Common pricing in various applications}
Finally, an extensive literature has studied the efficiency of common pricing policy, i.e. pricing without personalization, and how to design an efficient common pricing policy under practical applications.
\cite{common:banerjee} address the problem of designing pricing strategy for ride-sharing platforms, and reveal that under a reasonable condition, dynamic pricing that varies over time does not outperform fixed pricing. In contrary, \cite{common:castillo} show that fixed pricing \note{suffers} from a problem called {\em wild goose chase}, while the dynamic pricing solves it.
\cite{common:tong} consider the problem of spatial crowdsourcing and show that there exists an efficient dynamic pricing policy that significantly out-performs common pricing policy.
In context of cloud pricing, \cite{common:zheng} study the problem of computing an optimal profit-maximizing fixed price for cloud virtual service provider. The authors in \cite{common:abhishek, common:dierks, common:zheng2, common:song} consider the spot price exploited in Amazon Web Service, and study the profit of platforms and users under intersectional scenario where fixed pricing and spot pricing exist.
Likewise, we study the efficiency of common pricing for microtask crowdsourcing platform, and provide the optimal structures of common pricing policies for some canonical regimes.

%% file: model.tex
\section{Model and Problem Formulation}
\label{sec:model}

\subsection{System Model}
\label{sec:model1}
We consider a set $\set{N} = \{1,2, \ldots, n\}$ of workers, where $n$ is the number of workers who are available for a target task requested by the  requester.
Each worker $i \in \set{N}$ is associated with a {\em private} profile $(s_i, c_i) \in \mathbb{R}_{\geq 0}^2$: she produces the output of quality $s_i$ at cost $c_i$ if she decides to perform the task, where $s_i$ and $c_i$ quantify the individual contribution to the task and the opportunity cost to perform the task, respectively.
We adopt a standard Bayesian setting \cite{qualityaware:Han16}: the requester is aware of the prior distribution $f_i(s_i,c_i)$ for each $i \in \set{N}$.
We denote $x_i = 1$ if worker $i$ decides to work on the task and $x_i=0$ otherwise. Let $\vec{s} := [s_i]_{i \in \set{N}}$ and $\vec{c} := [c_i]_{i \in \set{N}}$. Then, the task requester has utility $u(\vec{s}\circ \vec{x})$ for some function $u:\mathbb{R}_{\geq 0}^n \mapsto \mathbb{R}_{\geq 0}$, where
$\vec{s}\circ \vec{x} = [s_i x_i]_{i \in \set{N}},$ a function of the collection of the qualities of task-accepting workers.

\paragraph{Quality-based pricing}
We consider a variant of budget feasible posted pricing framework where a {\em quality-based pricing policy} is posted {\em in advance} of workers' arrival
\footnote{We describe our pricing mechanism to be {\em posted pricing} since the pricing policy
is published a priori and each worker can accurately estimate the amount of payment as in the context of \cite{Posted:Balkanski16}. Though, our mechanism is not exactly on the same context as the ones used in the literature since the final payment to each worker depends on her private quality.}.
Note that it is often called as performance-based pricing in the literature~\cite{discrimination:shapiro}.
In quality-based pricing, an extra payment is made as bonus to workers who produce good-quality outputs in addition to a base payment which is paid to workers irrelevant to their quality.
We denote by $p:=[(p_i)]$ a pricing policy, where each worker $i \in \set{N}$ is offered $p_i$ consisting of an increasing function $p_i:\mathbb{R}_{\geq 0} \mapsto \mathbb{R}_{\geq 0}$ so that if worker $i$ accepts $p_i$ and submits an output $s_i$,  she will be paid $p_i(s_i)$.
To interpretational simplicity, we refer to the constant term of $p_i$ as a base payment, and the others as a bonus payment.
We assume that the workers 
strategically decide whether or \note{not to accept the task~\cite{behavioral:easley,behavioral:ghosh}}.
Then, the requester's objective is to find an optimal pricing policy $p$ that maximizes the expected utility for a given prior distribution $[f_i]$ of worker profile, while the choice of pricing policies can be restricted due to some practical constraints.
We remark that incentivizing a worker based on her quality could be either straightforward or requiring some additional process depending on the task.
For example in crowdsensing task, the requester might naturally have a method to estimate the quality of submitted sensing data. On the other hand, the requester can indirectly estimate $s_i$ by embedding golden questions with known answers purely for estimation, or delaying payment until the estimation on quality via some inference algorithms becomes accurate~\cite{Inference:1, Inference:2, Inference:3}, while allocating tasks to a conservative number of workers. For example in case of binary classification task, it is known that the classification error rate of belief propagation decreases exponentially on the number of workers per task, under a reasonable regime 
of fairly good workers whose average probability to provide correct response
is greater than $0.5$~\cite{Inference:4,Inference:5}.
In most cases, task requesters anyway perform a denoising process to obtain clean labels, in which worker quality can be simply assessed by comparison between the denoised labels and worker's responses, or a part of the denoising process.



\paragraph{Personalized and common pricing}
We classify pricing policies into two types: {\em (i) personalized pricing} and {\em (ii) common pricing}, depending on whether pricing is discriminatory between each worker or not. In personalized pricing, different pricing policy $p_i$ can be applied for different \note{worker} $i,$ while common pricing is the one which applies the same pricing function $p$ to all workers. Thus, we denote a {\nonprice} $p$ to be one with $p_i = p$ for each $i \in \set{N}$. Note that the set of all common pricing policies is a subset of all personalized pricing policies. 
Despite personalized pricing policy's more controllability on workers' behavior and thus more expected utility to the requester, 
{\nonprice} seems more useful, since {\perprice} is often discouraged in practice. For example, Amazon Mechanical-Turk \cite{Mturk:1} utilizes a common pricing. Also, even after the price to be posted is determined, {\perprice} still needs to observe each incoming worker to offer the customized price, leading to higher system complexity and privacy concerns which possibly hinder the workers from participation \cite{Discrimination:Anderson10}. However, {\nonprice} needs only a summarizing property of the entire worker profiles rather than individual ones, implying that {\nonprice} requires no information on each incoming worker after the price to be posted is determined.
Note that even under common pricing a certain form of price differentiation is played as an incentive, depending on how good each worker is. However,  
such a payment function is not differentiated across the workers, which is the key difference from personalized pricing.

\paragraph{Worker behavior} 
Given a pricing policy $p$ and worker profile $\vec{s}$ and $\vec{c}$, we assume that each worker is rational so that worker $i$ strategically maximize her quasi-linear payoff $p_i(s_i) - c_i$. More formally, worker $i$'s strategic decision on the task participation is made by:
\begin{align} \label{eq:worker_util} 
\phi_i{(p_i)} \defeq \mathbbm{1}[p_i(s_i) \geq c_i], 
\end{align}
where $\mathbbm{1}[A]$ is the indicator function of $A,$ i.e., $\mathbbm{1}[A]$ is $1$ if $A$ is true, and $0$ otherwise, i.e., worker $i$ accepts the task when $\phi_i = 1,$ and rejects otherwise. However, the final task allocation is determined depending on the available budget and the order of worker arrivals, as explained in what follows.
We assume that workers arrive in a given but latent ordering $\sigma$ drawn over all possible orderings of $\set{N}$. We note that our main results are based on the worst case analysis with respect to the ordering.
Then, for given budget $B$, arrival model $\sigma$ and pricing policy $p$, task allocation vector $\vec{x}$ is determined in the following manner, as in \cite{Posted:Balkanski16}:

\medskip
\begin{compactenum}[1)]
\item Worker $i$ arrives with profile $(s_i,c_i)$ with respect to ordering $\sigma$. 
\item If the remaining budget is smaller than worker \note{$i$'s} maximal possible reward, then discard the worker.
\item Else, offer her $p_i(s_i)$ and she accepts if her actual reward $p_i(s_i)$ is larger than or equal to her cost $c_i$, and get paid $p_i(s_i)$ and mechanism deducts it from the remaining budget. Otherwise, discard worker $i$.
\end{compactenum}
\medskip

Let $\set{P}(p, B, \sigma)$ be the (random) process generating a task allocation vector $\vec{x}$ under pricing policy $p$ and budget constraint $B$, where we denote $\vec{x}\sim\set{P}(p, B, \sigma)$. Note that the randomness in $\bm{x}$ comes from that of $\bm{s}$ and $\bm{c}.$
Under process $\set{P}(p, B, \sigma)$, after the entire budget is exhausted,
any incoming worker $i$ is naturally discarded and has $x_i = 0$ even if she wants to participate in the task, i.e., $\phi_i(p_i) =1.$ 
Hence, any realization $\vec{x}$ verifies 
the {\em ex-post} budget constraint \cite{Posted:Balkanski16}, i.e.,
\begin{align}
   \sum_{i \in \set{N}}p_i(s_i)x_i\leq B \;.
   \label{eq:budget_const}
\end{align}
We assume that $B = \Theta(n)$ to avoid the trivial arguments.
For given pricing policy $p$ and budget $B$,
we define the following expected utility (in the ex-post setting):
\begin{align}
U(B;p):=\EXP_{\vec{x}\sim\set{P}(p,B, \sigma)}[u(\vec{s} \circ \vec{x} )].
\label{eq:ex-post-util} 
\end{align}

\subsection{Problem Formulation}
For given $B \ge 0$, the problem of finding an optimal personalized pricing, referred to as {\perpriceprob}, is defined by: 

\vspace{-0.2cm}
\vspace{-0.2cm}
\begin{align}
\text{{\bf \perpriceprob}$(B)$:} \quad &
\underset{p=[p_i]}{\textnormal{max}} \ 
U(B;p) \label{eq:opt_goal}
\end{align}
In general, we call an instance of the above problem with different parameter 
an {\em optimal personalized pricing problem} throughout this paper.
Our goal is to investigate how feasible it is to employ a simple common pricing instead of an optimal personalized pricing. As earlier discussed, 
the payment function can be personalized to each worker's profile, while a common pricing $p_i = p$ for $i \in \set{N}$ is forced to be indistinguishable for all workers. Due to such controllability difference, the optimal {\perprice} naturally outperforms any optimal {\nonprice}.

In both personalized and common pricing, the following
tensions exist: If the requester sets high base and low bonus, it may fail
to recruit high-quality users and utilize only low-cost users. If
bonus is set too high with low base, it should spend a huge amount of
bonus on high quality users, so that the total number of recruited
users shrinks. Thus, it is necessary to strike a good balance between
base and bonus payments while satisfying the budget constraint to
maximize the achieved utility.

%% file: result.tex

\section{Main Results}
\label{sec:poa}
In this section, we present our main results which are summarized as the following two key messages: (i) a simple common pricing can approximate an optimal personalized pricing well, when an appropriate bonus mechanism becomes available and (ii) without bonus payment, no common pricing can achieve an reasonable approximate to optimal personalized pricing. In addition, we provide some useful insights on the effective structure of pricing policy with respect to the characterization of the crowdsourcing task.

\paragraph{Additive utility}
In our analytical results, we consider a canonical class of utility functions: additive mainly for mathematical tractability.
We say that $u(\cdot)$ is {\em additive} if
\begin{align}
\label{eq:utility_zero}
u(\vec{s} \circ \vec{x}) = \sum_{i \in \set{N}} s_i x_i .
\end{align}
Although our theoretical understanding in this section assumes additivity mainly for tractability,
we provide the experimental results which suggest similar implications for other utility functions.
It is worth to note that 
even with additive utility, finding the optimal personalized pricing is 
computationally challenging:
\begin{theorem}
\label{thm:hardness}
{\perpriceprob} with additive utility is NP-hard.
\end{theorem}
The detailed proof is provided in Appendix,
where we reduce classical knapsack problem, a well-known NP-hard problem, to a special case of {\perpriceprob}.

\paragraph{Reasonable workers}
For analytical tractability, we assume that workers 
have reasonable value at their labor
so that their quality given cost is upper bounded by a constant, formally the maximal value of the support of the random variable $s_i / c_i$ is constant, i.e. for every $i \in \set{N}$, $\prob{s_i / c_i < C} = 1$ for some constant $C$.
The assumption is very true in practice
due to human's limited capacity.
In addition, 
without this assumption, 
a simple policy 
hiring only unreasonably low-cost workers with unbounded $s_i / c_i$
easily achieves a certain level of utility
and thus devising efficient policy would not be meaningful in practice. 


\subsection{Price of Agnosticism and Power of Bonus}
\label{sec:poa1}
We now present our result that there exists a common pricing scheme whose performance is not far from that of an optimal personalized pricing. Our results are two-fold. 
First, we prove that there exists a simple, common pricing which produces a good approximation of OPP (small {\em price of agnosticism}, see Theorem~\ref{thm:ex_post}). 
Second, we show that there always exists a worker distribution and worker arrival under which any common pricing without bonus perform poorly (large {\em power of bonus}, see Theorem~\ref{thm:pad_bonus}).

\paragraph{$(\delta,\gamma)$-Approximation}
To formally discuss, we first define a notion of {\em approximate solution} given by a common pricing to OPP(B):

\begin{definition} \label{def:approximate}
For a given budget $B\ge 0$ and worker distribution $[f_i]$, consider an optimal personalized pricing problem $\text{\em \perpriceprob}(B)$ in \eqref{eq:opt_goal}, 
and let $p^\star(B)$ 
be an optimal personalized pricing of $\text{\em \perpriceprob}(B)$. 
For  $0 \le \delta \leq 1$, and $1 \leq \gamma$,
a common pricing $p_c$ is $(\delta,\gamma)$-approximate to {\em \perpriceprob}$(B)$ if
\begin{align*}
{U}(\gamma B;p_c) ~\geq~ \delta \times {U} \left(B; p^\star(B) \right).
\end{align*}
\end{definition}

The constants $\delta$ and $\gamma$ assess the suboptimality in terms of utility and budget, respectively. 
Intuitively, $\linear$ is able to 
achieve the utility of $\delta \times$ (utility by an optimal personalized pricing with budget $B$) with  budget $\gamma B.$ Note that $(1, 1)$-approximation corresponds to the exact optimality.

\paragraph{Price of Agnosticism}
We first study Price-of-Agnosticism (PoA) that 
quantitatively compares personalized and common pricing policies, 
revealing how good a simple common pricing with bonus payment is, compared to the best personalized one. 

We present Theorem~\ref{thm:ex_post} which states the existence of a common pricing which is a good approximation of an optimal personalized pricing, i.e., small price-of-agnosticism, as the number of workers $n$ grows.

\smallskip
\begin{theorem}[Price-of-Agnosticism]
\label{thm:ex_post}
There exists a common pricing $p_c$ that is $\left(1-O(e^{-n^{1/6}}), O(1)\right)$-approximate to {\em \perpriceprob}$(B)$.
\end{theorem}

\smallskip


\note{
This theorem indicates that we can find a common pricing endowed with an appropriate bonus function that is a $(1, O(1))$-approximate to the optimal personalized pricing with bonus, which indeed implies that one can design an approximately optimal pricing policy without any explicit pricing differentiation across the workers. This is quite surprising since it even holds when the worker's profile distributions are somewhat heterogeneous. It mainly springs from our assumption on reasonable workers which bounds the heterogeneity of the workers so that we can guarantee the approximately optimal utility within a slightly augmented budget.
}

We briefly summarize the key challenges and the issues in analysis. 
First, the major challenge in the proof is due to the tight correlation between task allocation and ordering of worker arrivals which the ex-post budget constraint creates.
In order to bypass this challenge, we first consider a tractable yet perhaps artificial budget constraint, referred to as the {\em ex-ante} budget constraint~\cite{Posted:Balkanski16}. Then, we establish the approximation ratio of $p_c$ under such ex-ante constraint by comparing it with an offline oracle algorithm in a fractionally relaxed version.
Finally, we connect the ex-ante approximation ratio to the ex-post one using the concentration inequalities for contention resolution scheme under the knapsack constraint~\cite{proof:Vondrak11}.

Specifically, we consider a class of {\em linear pricing} policy, i.e. $p(s_i) = x+ys_i$ for some parameter $x,y \in \mathbb{R}$. Efficiency of linear pricing is quite intuitive since linear pricing naturally recruits only {\em efficient} workers whose quality-to-cost ratio $s_i/c_i$ is at least some threshold, resulting in a cost-effective allocation of the budget in additive utility case. We note that linear pricing is revisited along with other pricing policies in Section~\ref{sec:simulation}.

We note that the offline oracle algorithm in the fractionally relaxed version actually outperforms OPP, where it is assumed to have the knowledge on the incoming worker's realizations of cost and quality \note{along with their arrival order} in advance, and also more controllability on the allocation vector. Thus, we conjecture that the actual performance between $p_c$ and OPP would be closer, and hence the gap between OPP and optimal common pricing would be much closer.

\paragraph{Power of Bonus}
We next study the role of bonus payment in common pricing. 
Intuitively, a common pricing without bonus has no ability to prevent spammers
from cherry picking: taking the (base) payment for low quality job.
In the following theorem, we demonstrate how vulnerable 
the common pricing without bonus payment can be to cherry picking.
\smallskip
\begin{theorem}[Power of Bonus]
\label{thm:pad_bonus}
There exists a worker profile distribution $[f_i]$ and worker arrival $\sigma$ such that no common pricing without bonus is $(\delta, o(n))$-approximate to 
{\em \perpriceprob}($B$) for any $\delta \in [0,1]$.
\end{theorem}

\note{Hence, without the bonus structure in pricing policy nor personalization, there always exists a problem instance that the requester suffers from any constant amount of utility loss even with budget augmented upto any $n^{1-\varepsilon}$ for any $\varepsilon > 0$.}
The sketch of the proof is as follows.
To obtain this arbitrarily worst-case result, we mainly consider a ``spammer-hammer" scenario in terms of worker qualities under the same cost (i.e., $c=c_i$ for all worker $i$), where we can always construct any common pricing without bonus that has arbitrarily poor performance even with additional budget by factor $o(n)$. This is intuitive, because as the quality gap between spammer and hammer increases, the need of price discrimination such as bonus grows accordingly.
This also implies that employing bonus payment is crucial for obtaining good approximation of common pricing to the optimal personalized pricing that is even computationally intractable. 

We note that the benefit of bonus can differ depending on
information about worker profile.
First, when we know each worker's profile exactly,
it is possible to design an optimal personalized pricing without bonus, which offers the minimal payment to engage workers.
Hence, if individual worker's profile and personalized pricing are available (although both of them are typically unavailable in practice), there is no need to use bonus payment.
On the other hand, in the case that the profile distributions are not deterministic but the workers are homogeneous, i.e. $\vec{s}, \vec{c}$ are identical for all workers, it is obvious that there is no gain at all from differentiating pricing policy between workers.
Meanwhile, bonus payments might give some amounts of gain, we highlight that the power of bonus is larger than that of personalization in this case.

\subsection{Profile Distribution and Optimal Common Pricing}
\label{sec:opt}
We now study what an effective structure of bonus function for common pricing is, 
in three regimes of profile distribution.
Each regime corresponds to different type of tasks. 


\paragraph{Ratio-equivalent regime and linear pricing}
Consider tedious tasks such as filtering out the spam mails, correcting the typos on some elementary-level articles, or classifying the animals from images. The worker quality is linearly proportional to her effort and time.
Hence, the ratio between the cost and quality would be the same across the workers.
Formally, we define {\em ratio-equivalent} regime 
such that $\prob{c_i = k s_i}=1$ for every $i \in \set{N}$ and for proportional constant $k \in \mathbb{R}_{\geq 0}$.
We next define a {\em linear-pricing} with parameter $\alpha$ as the following:
\begin{align*}
    \text{linear-pricing:}\quad p_{lin}(\alpha) = \alpha s_i
\end{align*}
Note that it is a common pricing policy regardless of the parameter $\alpha$.
We find that $p_{lin}(\cdot)$ with proper parameter exactly achieves the optimality in this case as follow:
\begin{proposition}\label{prop:1}
Under ratio-equivalent regime with proportional constant $k$, linear-pricing $p_{lin}(k)$ is asymptotically optimal, meaning that 
$p_{lin}(k)$ achieves $(1,1)-$approximate to OPP(B) as $n$ goes to infinity.
\end{proposition}
The requester's fundamental goal  can be translated into recruiting the workers with higher marginal utility-to-cost ratio, i.e. $s_i/c_i$. Since the linear bonus function naturally achieves this goal in additive utility function, by recruiting the workers in the descending order of $s_i/c_i$, we expect that linear bonus equipped with a proper base payment would behave well in practice.

\paragraph{Cost-equivalent regime and threshold pricing}
SETI \cite{ex:SETI}
crowdsources computation resource,
where the only thing participant needs to do is just setting
a specific screen saver on computer,
and thus the effort to participate is almost identical across the entire workers
while each worker's contribution is different depending on the specification of computers.
This motivates us to define {\em cost-equivalent} regime such that $\prob{c_i = c_0} = 1$ for every $i \in \set{N}$ and for some $c_0 \in \mathbb{R}_{+}$.
For this regime, we consider the following form of common pricing with parameters $\alpha,\beta$:
\begin{align*}
    \text{threshold-pricing:}\quad p_{thre}(\alpha, \beta) = \alpha \mathbbm{1}[s_i \geq \beta]
\end{align*}
In here, $\alpha$ denotes the amount of bonus payment, and $\beta$ denotes the threshold of worker quality to grant the bonus payment, where $\beta$ controls for whom we need to grant the bonus, and
$\alpha$ controls how much to motivate those qualified workers.
The following proposition implies that under the cost-equivalent regime, $p_{thre}(\cdot)$ with proper parameter is optimal:
\begin{proposition}\label{prop:2}
Under cost-equivalent regime, threshold-pricing $p_{thre}(c_0, \beta_B)$ is asymptotically optimal for $\beta_B = \sup \left\{ \beta \geq 0 : 
\sum_{i \in \set{N}} \prob{s_i \geq \beta} \geq B/c \right\}$.
\end{proposition}
This is quite intuitive since in cost-equivalent regime, any optimal pricing policy needs to offer exactly the same price to every worker and hires the worker in the order of higher quality, where threshold pricing naturally achieves it.

\paragraph{Quality-equivalent regime and effect of bonus}
We define {\em quality-equivalent} regime such that  $\prob{s_i = s_0} = 1$ for every $i \in \set{N}$ and for some $s_0 \in \mathbb{R}_{\geq 0}$.
As an example corresponding to this regime, one
can consider survey asking non-trivial question, e.g., Moral Machine~\cite{ex:MoralMachine} collecting people's decision on a moral dilemma,
where
each individual's contribution is counted as just a single answer
regardless of the depth of thinking which may differ person-by-person significantly.
We provide the following proposition which implies we can always transform any common pricing into common pricing without bonus with the same expected utility.
\begin{proposition}\label{prop:3}
Under quality-equivalent regime, for any common pricing $[p(s_i)]$, there exists a common pricing without bonus $[(p'(s_i))]$ such that $p'(s_i) = \nu$ for some $\nu \in \mathbb{R}_{\geq 0}$ and $U(B; p) = U(B, p')$.
\end{proposition}
If there exists no difference in worker quality, the requester just need to recruit any workers in the cheapest manner and this can be done without any bonus payment, which implies that there exists no gain from exploiting bonus payment in this regime.
Note that we're not claiming that bonus payment harm the requester's utility, but there is no need of adopting bonus payment.


%% file: simulation.tex
\section{Simulation}
\label{sec:simulation}
In this section, we numerically analyze the utility of various {\em common} pricing along with in terms of their PoAs and PoBs, and verify the structure of efficient common pricing structure with respect to the profile distributions.

\subsection{Setup}
We consider 100 workers and two regimes of cost distributions of worker cost $c_i$, where for Figure~\ref{fig:sim}\subref{fig:sim_linear} and~\ref{fig:sim}\subref{fig:sim_affine}, cost of each worker $i$ is drawn uniformly at random from $[0.3, 0.7]$, and for Figure~\ref{fig:sim}\subref{fig:sim_uniform}, cost of every worker $i$ is $0.5$ in deterministic manner.
For each figures, we consider a mixture of two types, spammer and hammer, of worker quality where {\bf Spammer} is a worker with $s_i = 0.1$, and {\bf Hammer} consists of the following 3 sub-types of worker's quality distribution: {\bf (Linear)} $s_i = c_i$, {\bf (Affine)} $s_i = c_i - 0.1,$ and {\bf (Uniform)} $s_i \sim U[0.3,0.7]$, i.e. uniformly randomly drawn over $[0.3, 0.7]$.
The budget is set to be $30,$ and we choose the additive utility function of the requester $u(\vec{s} \circ \vec{x}) = \sum_{i \in \set{N}} s_i x_i .$ Depending on the type of bonus payment and the value of base payment, we consider different common pricing schemes. 
First, {\bf $x$-linear} pricing represents pricing policy form of $p(s_i) = x+ks_i$ for some bonus constant $k$.
Next, {\bf $50\%$-thre(shold)} pricing is defined as $p(s_i) = k\mathbbm{1}[s_i \geq 0.5]$ so that it incentivizes the worker only if her quality is at least some threshold for some bonus constant $k$.
For both pricing policies, we choose the bonus constant $k$ so that the expected utility becomes the largest over every pricing policy with the corresponding form.
{\bf No-bonus} pricing is the optimal one among the common pricing without bonus payment. Finally, we compute the expected utility of OPP\footnote{Actually, we compute the expected utility of offline problem, where it is possible to lookahead the realized profile of the workers, and pay them the exact cost. We note that it is exactly the same with classical 0-1 knapsack problem} in a brute-force manner, which was possible thanks to the constrained form of profile distributions and a reasonable number of workers.
Finally, in Figure~\ref{fig:sim}\subref{fig:sim_linear} all the workers are of the types of spammer or linear, where the proportion of linear workers increase from $0\%$ to $100\%$,  Figure~\ref{fig:sim}\subref{fig:sim_affine} corresponds to the case of spammer and affine workers, and Figure~\ref{fig:sim}\subref{fig:sim_uniform} corresponds to the case of spammer and uniform workers with equivalent cost.

\begin{figure}[!t]
\centering
\subfloat[]{\includegraphics[width=0.33\columnwidth]{fig/sim_linear.eps}
\label{fig:sim_linear}
}
\subfloat[]{
\includegraphics[width=0.33\columnwidth]{fig/sim_affine.eps}
\label{fig:sim_affine}
}
\subfloat[]{
\includegraphics[width=0.33\columnwidth]{fig/sim_uniform.eps}
\label{fig:sim_uniform}
}
\caption{{\em (a)} Expected utility of pricing policies with respect to the proportion of linear workers in spammer-linear scenario. $0\%$ means that all the workers are spammers. {\em (b)} Expected utility of pricing policies with respect to the proportion of affine workers in spammer-affine scenario.{\em (c)} Expected utility of pricing policies with respect to the proportion of uniform workers in cost-equivalent spammer-uniform scenario.
}
\label{fig:sim}
\end{figure}


\subsection{Evaluation}
\paragraph{Price of agnosticism and power of bonus}
In Figure~\ref{fig:sim}\subref{fig:sim_linear} and~\ref{fig:sim}\subref{fig:sim_affine}, we observe that linear pricing has a good match with OPP, and in Figure~\ref{fig:sim}\subref{fig:sim_uniform}, threshold pricing achieves almost the optimum without any auxiliary budget. It implies that price of agnosticism is small when the pricing policy is appropriately designed, as we analyzed in Section~\ref{thm:ex_post}.
Next, we observe that the expected utility of No-bonus is \note{very} small, compared to 
those with bonus for both scenarios, where it supports the importance of bonus, as stated in Theorem~\ref{thm:pad_bonus}.
Finally, we highlight that in order to utilize the power of incentivization, it's important to design the bonus function properly with respect to the profile distributions, since threshold pricing performs even worse than No-bonus for spammer regime in Figure~\ref{fig:sim}\subref{fig:sim_linear}.

\paragraph{Choice of base payment}
In Figure~\ref{fig:sim}\subref{fig:sim_affine}, it is interesting to see that pricing policies with base $0.1$ have larger expected utility than 
that with zero base payment. 
For a linear pricing to achieve near-optimal utility without any augmented budget, it needs to reduce its budget waste caused by redundant bonus payment. To this end,  every worker needs to output exactly proportional to what the requester paid, i.e., outputs $0.2$ if she gets paid $\$0.2$, outputs $0.4$ for $\$0.4$.
For affine workers, $0$-linear fails to do so due to the negative bias introduced in the affine profile distribution. For example, if $0$-linear pays $2\cdot s$ bonus, then it pays $\$0.4$ for worker with profile $(s,c) = (0.2,0.4)$, but it pays $\$0.8$ for worker with profile $(s,c) = (0.4, 0.6)$, i.e., the budget is wasted on this worker which leads to the utility decrement as shown in Figure~\ref{fig:sim}\subref{fig:sim_linear}.
However, if some amount of the base payment is ready, it helps to recover the negative bias in affine workers, so that they output exactly as they get paid.

\paragraph{Efficient bonus structure}
In the right-end side of Figure~\ref{fig:sim}\subref{fig:sim_linear}, we observe that linear pricing achieves near-optimal utility as we claim in Proposition~\ref{prop:1}. We also find that linear pricing is robust to the spammer since the existence of spammers does not significantly lower the performance of linear pricing compared to OPP.
In the right-end side of Figure~\ref{fig:sim}\subref{fig:sim_uniform}, where it indicates the cost-equivalent regime, we find that threshold pricing achieves the near-optimal utility as we observe in Proposition~\ref{prop:2}. We note that threshold pricing works well in the purely uniform worker regime, but it is worse than linear pricing and even pricing without bonus, as the proportion of the spammer increases. Hence, we conclude that threshold pricing works well if it is designed well upon the specified regime, but is not robust under the various profile distribution regimes.
Finally, in left-end side of Figure~\ref{fig:sim}, we observe that pricing policy with bonus does not outperform the pricing without bonus in significant manner. We note that this regime refers to the quality-equivalent regime since all the spammer has the same quality, and it aligns with Proposition~\ref{prop:3}.
We finally note that the linear and threshold pricing would be revisited in real-world experiment with additive and non-additive utilities in the next section.

%% file: experiment.tex
\section{Real-world Experiment}
\label{sec:experiment}
In this section, we provide real-world experiments carried out at  Amazon Mechanical Turk (Mturk) to verify our findings on efficiency of common pricing, and investigate how common pricing policies perform for both additive and non-additive utility functions. 

\subsection{Setup}
\label{sec:experiment-setup}
We requested workers to correct a typographical error, i.e. typo, in an article of $426$ words with $15$ typos, where 
we choose an article from CNN~\cite{CNNarticle} and alter some words into hand-made typos so that  
these typos are almost evenly distributed.
We consider two utility functions: {\bf (U1)} the total number of typo corrections done by workers (Figure~\ref{fig:exp}\subref{fig:exp-util}) and {\bf (U2)} the number of corrected typos (Figure~\ref{fig:exp}\subref{fig:apd_utility}), where we assume that a typo is corrected if a suitable number $n_\text{cor}$ of workers correct the typo. In our case, we choose to be $n_\text{cor}=8$. Note that it is not hard to see that {\bf U1} is additive but {\bf U2} is non-additive. 

\begin{figure}[!h]
\centering
\subfloat[]{\includegraphics[width=0.49\columnwidth]{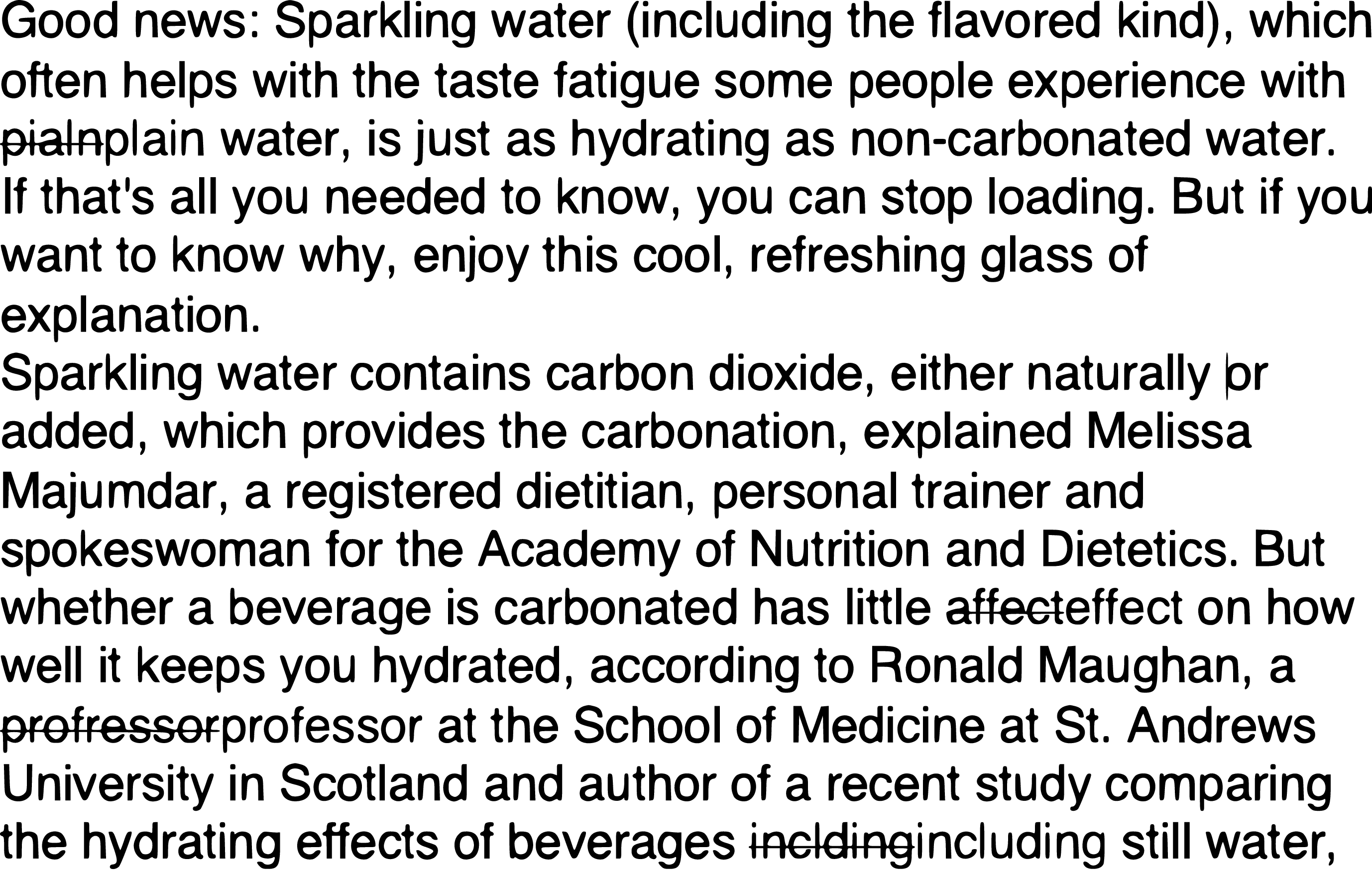}
\label{fig:task}
}
\subfloat[]{
\includegraphics[width=0.5\columnwidth]{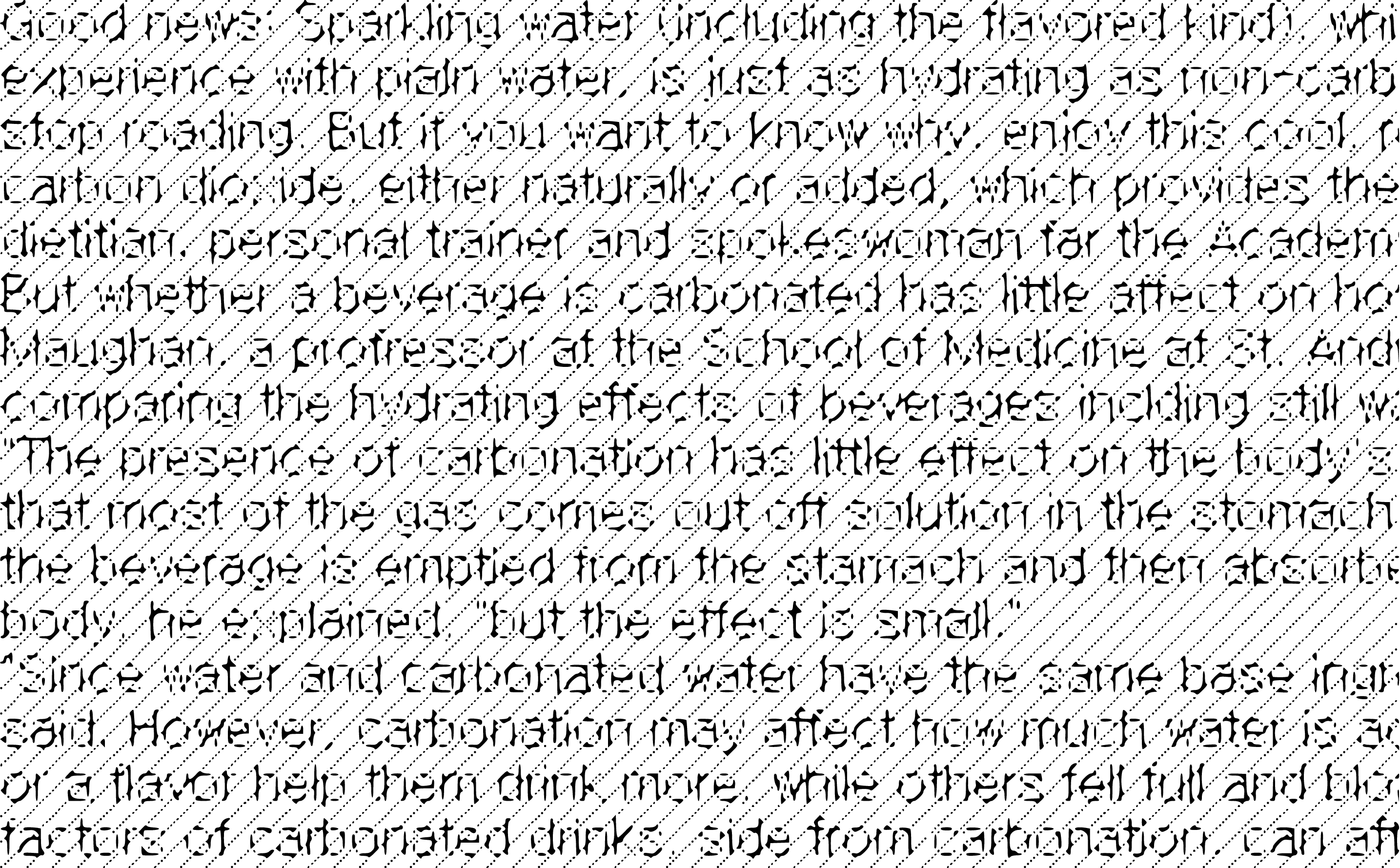}
\label{fig:task-display}
}
\vspace{-0.3cm}
\caption{(a) The typo correction task used in the experiment, where typos highlighted in red are inserted
    instead of original words shown in blue; and
    (b) The actual display shown to workers with dashed lines to prevent to use of automated tools.
}
\label{fig:exp-task}
\vspace{-0.3cm}
\end{figure}


We consider the following four common pricing schemes (where base and bonus payments are in the unit of USD):
\smallskip
\begin{compactitem}[$\circ$]
\item {\bf $b$-base:} This gives the base payment $b=0.7$ without bonus.
\item {\bf linear:} With the base payment of 0.5, bonus is given proportional to the number of corrected typos with the maximum of $0.5,$ i.e.,  $0.5\times$(num. of corrected typos)/15.
\item {\bf $m\%$-thre(shold):} 
With the same base payment of 0.5, this additionally grants bonus $0.5$ to the worker who has corrected more than $m\%$ of the total $15$ typos, where $m=50,90.$ Note that this is the same as $(0.5, m/100)-$threshold pricing with base payment $0.5$ described in Section~\ref{sec:simulation}
\end{compactitem}
\smallskip

The {\bf $m\%$-thre} can be considered as a version of {\bf linear} that has the drastic increase of bonus at some point. We recruit 520 workers, where, 
in order to prevent duplicated worker, we separate them into different 8 groups\footnote{We also tried four additional pricing schemes, $b$-base ($b=$0.1, 0.3, 0.5, 0.9), not for evaluation of common pricing, but just for computing the simulated OPP.} at random, each of which correspond to one common pricing.


To examine the efficiency of common pricing schemes, it is necessary to compare them to an ideal one such as the optimal personalized pricing (OPP). However, Mturk {\em does not allow such a personalized pricing} as well as unavailability of each incoming user's worker profile. 
Thus, we compute the values of the following {\em simulated OPP} instead:
We first assume that the workers have deterministic profiles which are sampled from 
$s \sim N(xc+y, z)$ for some latent variables $x,y,$ and $z$ (i.i.d. across workers). We model this distribution using the data collected from {\bf $b$-base} policies where $b\in \{0.1,0.3, \ldots, 0.7\}$. Then, we divide $24$ hours into $65$ slots and allocate one worker for each slot with its profile sampled from our constructed joint distribution. 
Then we compute the latent variables $x,y$ and $z$ from the sample mean and sample standard deviation statistics, where we get $s \sim N(2c+7, 4)$.
Finally, we compute the worker allocation of OPP in a brute-force manner, given the budget and worker profile samples, then by using the allocation, we compute the expected utility for every slot where expectation is taken over repeating sampling of profiles.

\begin{figure*}[!t]
\centering
\subfloat[Additive utility: Num. of all typo corrections by workers 
]{\includegraphics[width=0.238\columnwidth]{fig/figB.eps}\label{fig:exp-util}}
\hspace{0.1cm}
\subfloat[Non-additive utility: Num. of corrected typos]{\includegraphics[width=0.235\columnwidth]{fig/figC.eps}\label{fig:apd_utility}}
\hspace{0.1cm}
\subfloat[Additive utility with respect to used budget]
{\includegraphics[width=0.238\columnwidth]{fig/uToB.eps}\label{fig:exp-uAToB}}
\hspace{0.1cm}
\subfloat[Non-additive utility with respect to used budget]
{\includegraphics[width=0.236\columnwidth]{fig/uToB_nonsep.eps}\label{fig:exp-uAToB-nonsep}}
\hspace{0.1cm}
\subfloat[Used budget]
{\includegraphics[width=0.2385\columnwidth]{fig/figD.eps}\label{fig:exp-budget}}
\hspace{0.1cm}
\subfloat[Num. of participants]{\includegraphics[width=0.2385\columnwidth]{fig/figG.eps}\label{fig:num-worker}}
\hspace{0.1cm}
\subfloat[Used budget by Num. of participants ]{\includegraphics[width=0.2365\columnwidth]{fig/figH.eps}\label{fig:num-worker-by-budget}}
\hspace{0.1cm}
\subfloat[CDF of worker quality (\# of correct answers)]{\includegraphics[width=0.242\columnwidth]{fig/figE.eps}\label{fig:cdf-worker}}
\vspace{-0.3cm}
\caption{Experimental results of a typo correction task in Amazon Mechanical Turk over $24$ hours with various pricing policies. In (b),  a typo is regarded as being corrected, if corrected by at least $8$ workers.} 
\label{fig:exp}
\vspace{-0.4cm}
\end{figure*}




\subsection{Evaluation}
\label{sec:eval}
\paragraph{Efficiency of common pricing}
Figure~\ref{fig:exp}\subref{fig:exp-util} shows the accumulated number of typo corrections, corresponding to an additive utility. 
In Figure~\ref{fig:exp}\subref{fig:exp-util}, we observe that the common pricing {\bf $50\%$-thre} and {\bf linear} nearly achieves up to $83\%$ and $91\%$ of the utility of OPP at $t = 24$, respectively. Figure~\ref{fig:exp}\subref{fig:exp-budget} represents the budget consumption of policies with respect to elapsed time. We observe that the total budget consumption for {\bf $50\%$-thre} and {\bf linear} are just about $1.4$ and $1.3$ times more than that of OPP. Combining these two results, we verify that with small additional  budget in common pricing with bonus policy, their achieved utility 
is close to that of OPP. Moreover, we note that the simulated OPP is designed to underestimate the overall cost distribution of workers, because any workers with cost lower than $0.7$ will accept the {\bf $0.7$-base} policy, but we just assume that their is $0.7$. It implies a smaller gap between {\bf $50\%$-thre} and {\bf linear} and the real OPP.


\paragraph{Power of bonus}
In Figures~\ref{fig:exp}\subref{fig:exp-util}-\ref{fig:exp}\subref{fig:exp-budget}, 
employing bonus significantly helps to improve the utility or save the budget spent. 
For example, {\bf $0.7$-base} spent almost $2$ times
more budget than {\bf $90\%$-thre}, but they collect a similar number of typo corrections at the end. This means that the bonus payment can double the ratio of utility to budget. This supports the non-negligible power of bonus as analyzed in Theorem~\ref{thm:pad_bonus}, where bonus payment incentivizes high quality workers and increases the efficiency of budget usage.
Indeed, in Figure~\ref{fig:exp}\subref{fig:cdf-worker},
the pricing without bonus, e.g., {\bf $0.7$-base,} shows to be the most skewed distribution towards poor workers due to the lack of incentive for quality workers.

\paragraph{Structure of utility and bonus}
We now study the impact of the form of utility function and bonus.
In Figure~\ref{fig:exp}\subref{fig:exp-util}, we observe that {\bf linear}  eventually achieves the highest utility among all the common pricing policies, and also the highest final utility-to-budget ratio in Figure~\ref{fig:exp-uAToB}. For additive {\bf U1}, to achieve high utility, the requester has to maximize $\sum_i s_i/B \sim \sum_i s_i/\sum_{i} c_i$. Hence the requester naturally needs to maximize $s_i/c_i$ for each worker $i$, and we find that {\bf linear} naturally 
achieves this behavior, since it only recruits workers with $s_i/c_i$ larger than or equal to a specific threshold. Therefore, if the requester wants to maximize an additive utility, (e.g., findings  many as possible typos in the typo-correction task), {\bf linear} would be a proper answer. Next, for a non-additive {\bf U2} in Figure~\ref{fig:exp}\subref{fig:apd_utility}, we observe that {\bf $50\%$-thre}  achieves the highest utility, and the highest final utility-to-budget ratio in Figure~\ref{fig:exp}\subref{fig:exp-uAToB-nonsep}. In this case, maximizing per-worker contribution-to-cost ratio does not necessarily optimize the utility. Hence, if the requester decides to trust the corrected typo, if at least $n_\text{cor}$ workers give the same answer, she needs to exploit {\bf $m\%$-thre}. However, we also find that if the bonus granting condition is too cruel as in {\bf $90\%$-thre}, the requester fails to exhaust given budget with $n$ workers (or within a fixed time), which leads to low utility as seen in Figures~\ref{fig:exp}\subref{fig:exp-util} and~\subref{fig:apd_utility}. We find that if the bonus granting condition is too critical, it could happen that no one is possible to earn the bonus regardless of the bonus amount, and hence only the base payment motivates the workers in this case.
This ``extreme'' bonus policy will help more when the requester has enough time, or equivalently has large pool of workers, but experience lack of budget, as we observe that the ratio of {\bf U1} (or {\bf U2}) to
consumed budget is the largest in {\bf $90\%$-thre}.

%% file: conclusion.tex
\section{Conclusion}
\label{sec:conclusion}
In this paper, we analytically studied the impact of personalization and bonus payment in posted pricing. We prove that a common pricing policy equipped with proper bonus nicely approximates the optimal personalized pricing, where bonus is inevitable for common pricing policy to work well. Our analytical findings, verified through simulations and real experiments, explain why many current crowdsourcing systems work well with simple pricing policies, and present some practical implications on how to price workers better with a simple mechanism according to the characteristic of the crowdsourcing tasks.
In what follows, we discuss few future research directions
regarding possible irrational behaviors of workers in practice.


\paragraph{Prospect theory}
In our model, we assume that each worker~$i$ knows the exact value of quality~$s_i$ and cost~$c_i$, and computes the best response. However, in practice, there can be some uncertainty in the self-evaluation and expected utility, and this makes
worker behavior can be more complex than what we assume.
In particular, once we introduce such an uncertainty, we may need to consider
risk-averse or risk-seeking behaviors in the literature of prospect theory~\cite{prospect:kahneman}, while workers in our model 
are basically assumed to be risk-neutral.
Indeed, the risk-averse behavior is partially observed in 
our experiment observing 
the less number of participants recruited by 
risky pricing (such as {\bf $50\%$-thre} or {\bf $90\%$-thre}) than that by 
safe pricing (such as {\bf linear} or {\bf $0.7$-base}) 
in Figure~\ref{fig:exp}\subref{fig:num-worker}, 
although our model can also explain this as well.
\note{In this sense, we believe that our empirical results might be utilized to reverse-engineer the crowdsourcing worker's utility in such platform in the context of behavioral game theory~\cite{behaviorgt:camerer,behaviorgt:crawford}.}
After all, since the risk-averse or risk-seeking natures 
are essentially irrational behavior of workers,
we believe the power of bonus in practice is greater than
what we studied in this work assuming risk-neutral workers,
while we need careful analysis and experiment to compare personalized and common pricings.

\paragraph{Assessing quality of contribution}
\note{
Our study suggests quality-based pricing for efficiency. 
Thanks to an extensive line of work \cite{Inference:1,Inference:2,Inference:3,Inference:4,Inference:5}, there are a number of 
off-the-shelf algorithms for worker assessment.
However, they could be erroneous in some scenarios. Imprecise worker assessment might discourage the worker's participation, especially those who tend to avoid a large variability in their reward. This might require us to construct more robust common pricing policy to encourage the participation of such workers.
Besides, there might be a strategic tendency to collude for the workers if their contribution is estimated based on the other's contribution in collective manner. In this context, one might seek to design a pricing policy along with truthful quality assessment process by integrating the concept of proper scoring rule in information elicitation literature~\cite{Eliciting:Ho16,Eliciting:Liu16,Eliciting:Radanovic16,Eliciting:Chen20}.
Nonetheless, we believe that the workers in microtask crowdsourcing platform tends to truthfully report their belief for the most tasks since the tasks are usually simple and easy so that each worker would think that the correctly submitted answers will be majority.
}

\paragraph{Mechanism complexity}
\note{Posted price mechanism indeed simplifies the communication between the requester and workers, and encourages the participation of volatil workers.
However, quality-based pricing introduces additional complexity in the worker's point of view, since each worker needs to estimate own quality in advance to determine whether or not to accept the task by quantifying the expected amount of reward.
Since it might discourage the volatile workers' participation, this is indeed an important factor which requires further study.
}
As an evidence, one can interpret
that in Figure~\ref{fig:exp}\subref{fig:num-worker}, 
{\bf linear} and {\bf $0.7$-base}
engaged more participants than
{\bf $50\%$-thre} and {\bf $90\%$-thre} due to their relative simplicity.
\note{In this regard, we might consider such mechanism complexity as an additional variable in our optimization problem, whether as an objective or constraint.
The line of works studying the menu-size complexity of auction mechanism~\cite{menu:babaioff,menu:chawla,menu:chen} could be a guide for such extension.
}
We however note that all the bonus payment schemes which we consider in this paper are much simpler than auction-based ones, e.g., \cite{complexity:einav2018}.

%% file: appendix.tex
\section{Experimental detail}

\begin{figure}[!t]
\centering
\subfloat[]{\includegraphics[width=0.45\columnwidth]{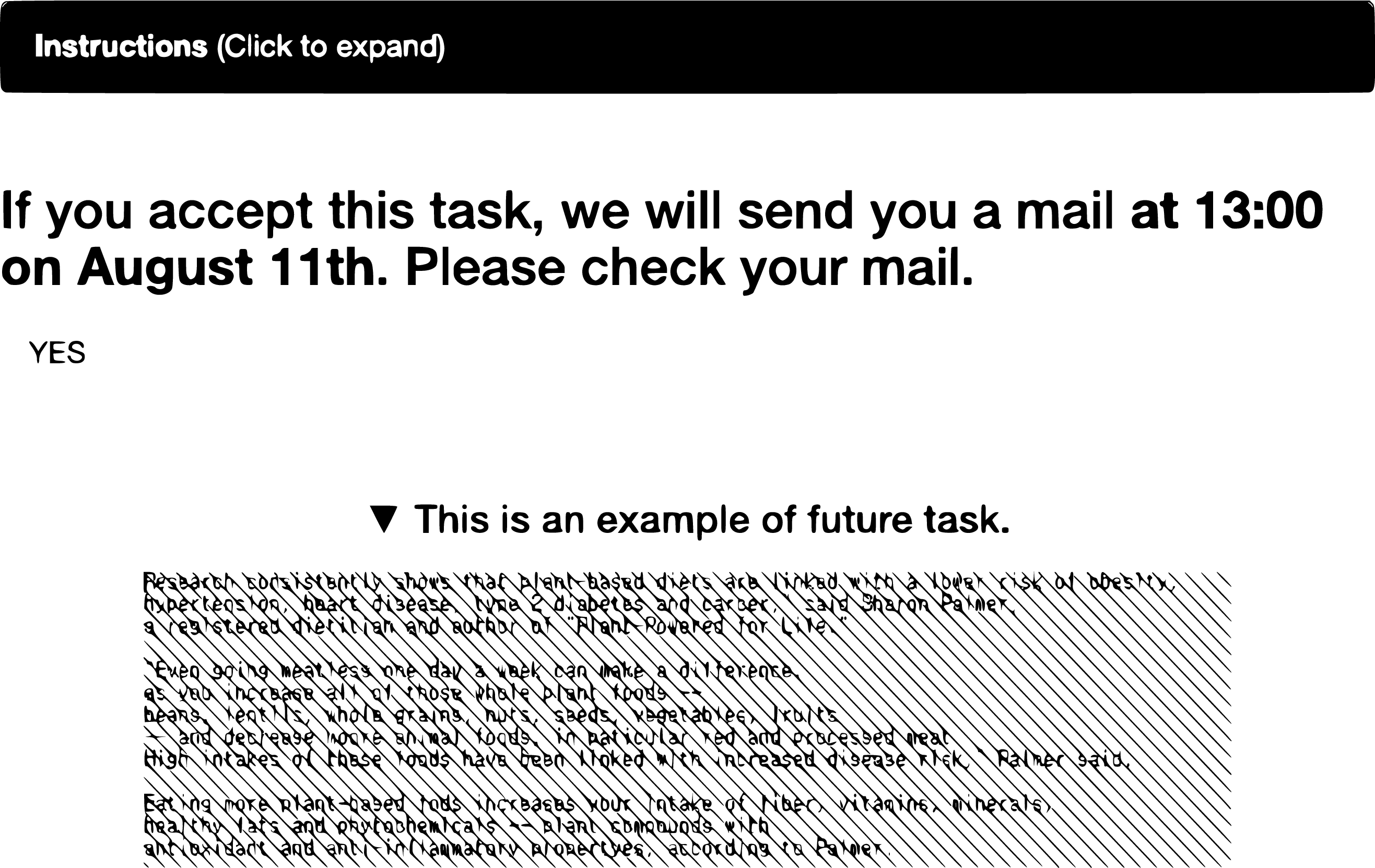}
\label{fig:pre_test}
}
\hspace{0.3cm}
\subfloat[]{
\includegraphics[width=0.42\columnwidth]{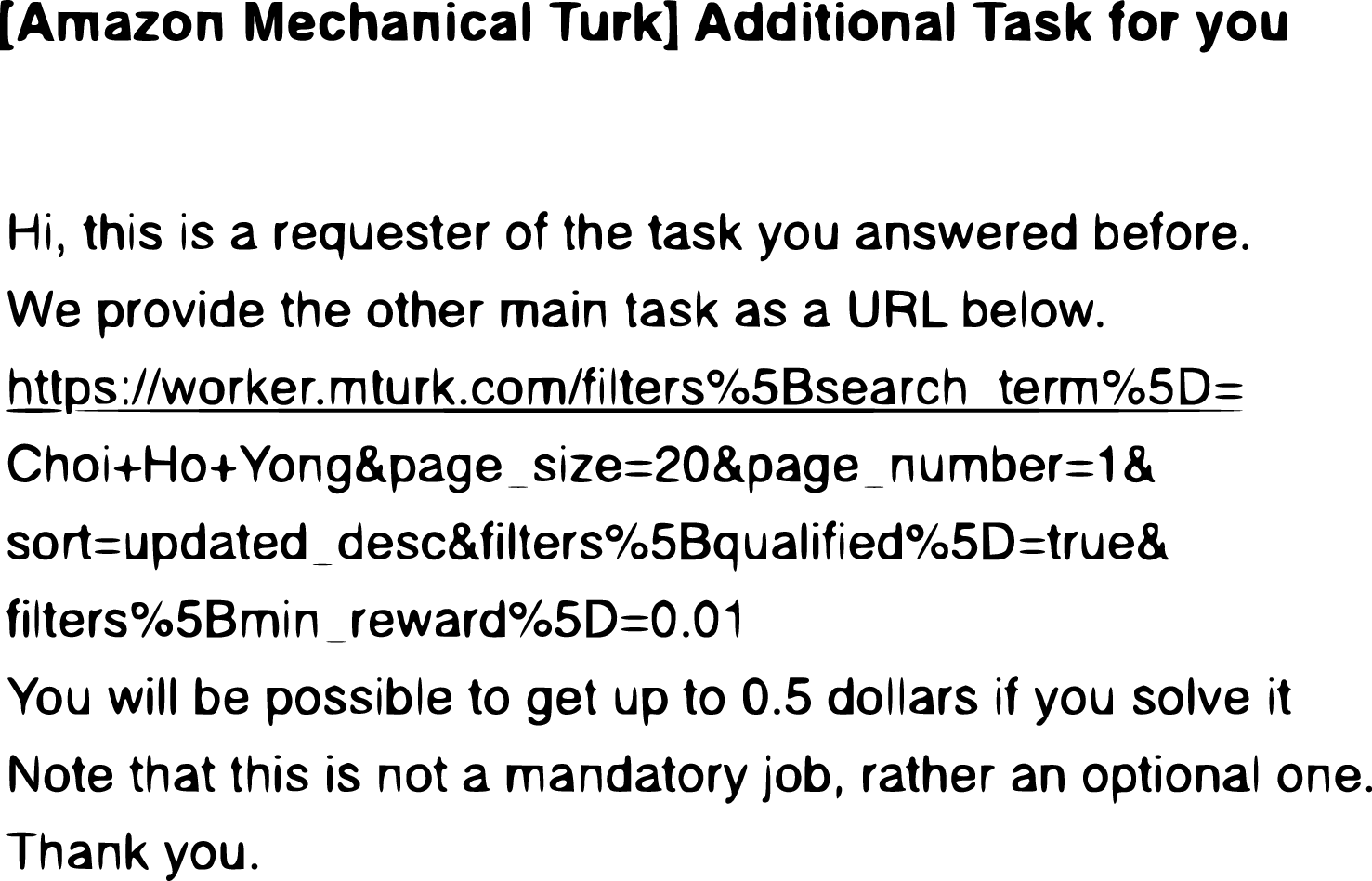}
\label{fig:e_mail}
}
\caption{{\em (a)} The actual pre-task display used in the experiment {\em (b)} The e-mail which was used to provide main task
}
\label{fig:exp_setup}
\end{figure}

\note{
We briefly present how we design our Mturk experiment to fairly compare the various pricing schemes.
To evenly allocate workers to various pricing schemes, we firstly post a preliminary task at $\$0.02$ to recruit the workers in advance.
In the task, we require the workers to agree on receiving an e-mail which asks the workers
to participate on a further main task at a specified time.
As presented in Figure~\ref{fig:exp_setup}\subref{fig:pre_test}, we attach an illustrative example for the main task, and the workers are committed to type "YES".
This process enables us to recruit $520$ workers within two days, and we randomly distribute them to each experimental group equipped with corresponding pricing scheme.
Each group is finally offered to participate on our main task as described in Figure~\ref{fig:exp_setup}\subref{fig:e_mail}, where the provided pricing contract differs across the groups.
We believe that we encourage their strategic decision by explicitly specifying that the task is optional and there would be no further disadvantage in not completing the main task.}

\note{In the main task, the workers are requested to find and correct typos as we briefly discussed.
The typos are almost evenly distributed over the article, and for each target word, we alter it into an elementary level of typo so that there exists no fundamental difficulty in correcting the typos whenever each  worker discovers it. 
When posting the task in Mturk, we intentionally coat dashed lines in the display of the article in order to prevent the workers to exploit some automated tools, like OCR, which might enable them to solve the task without exerting their own effort. To choose a proper price for our task, we presume that the most workers would spend about $5$ to $10$ minutes in completing the task, and decide the base and bonus payment so that the expected payment divided by spent time stays near in statutory minimum wages.}

\note{
We now provide the details of how we model and compute the performance of OPP in approximate manner, which we presented as {\em simulated OPP} in Section 5.
We first assume that the workers are sampled from i.i.d. joint distribution where $c_i \sim \text{Uniform}(\{0.1, 0.2, \ldots, 2.0\})$ and $s_i \sim \set{N}(xc_i+y, z^2)$, and they arrive with geometric distribution $G(\lambda)$ for some hidden parameters $x,y,z,\lambda>0$.
To infer $x$ and $y$, we compute expected quality for every {\bf $b$-base} policy where $b\in \{0.1,0.3,0.5,0.7\}$, and then use linear regression over $b$ and expected quality, which leads us to obtain $x=2,y=7,$ by assuming $b=c_i$\footnote{In this way, the quality distribution will be overestimated since the workers accepting {\bf $b$-base} policy will actually possess cost higher than $b$. Eventually, this leads to an overestimation of utility, and finally will result a conservative comparison between utility of OPP and common pricing policies, i.e. the actual utility gap might be much lower }. We obtain $z=4$ by computing the expectation of standard deviation for every {\bf $b$-base} policy. Finally, we get $\lambda=0.13$ by computing expected arrival interval between any two consecutive workers for every {\bf $b$-base} policy.
Given this model, we first assume that the total budget $B$ is $14$ as in the total consumed budget in $0.7$-base policy. Then, we sample $65$ data from the profile distribution and their arrival time from arrival model, and compute utility at each time of OPP computed from this sample. Finally, we repeat this process 100 times to estimate an expected performance over worker profile and arrival samples.
}

\section{Proof of Main Results}
\label{sec:appendix}

\subsection{Proof of Theorems}
\paragraph{Proof of Theorem~\ref{thm:hardness}}
To prove, we consider deterministic worker profiles, i.e., $\prob{({s}_i,{c}_i) = (\bar{s}_i,\bar{c}_i)~\forall i \in \set{N}}~=~1$ for some $[(\bar{s}_i, \bar{c}_i)]$. Since it is a special case of our problem, it is sufficient to show that {\perpriceprob} is NP-hard even on this simple setup.
Consider 
the following generalized 0-1 knapsack problem ({GKP}): for $\vec{s}, \vec{c} \in \mathbb{R}^n_{\geq 0}$,
\begin{subequations} \label{eq:GKP}
\begin{align}
\text{\hspace{-1cm}\bf GKP:} \quad 
\underset{\vec{x}\in \{0,1\}^n}{\textnormal{maximize}}  & \quad
u(\vec{{s}} \circ \vec{x} ) \nonumber  \\
  \textnormal{subject to} & \quad
  \sum_{i=1}^{n}{c}_ix_i\leq B \;. \label{eq:GKP-condi}
\end{align}
\end{subequations}
It is well-known that {GKP} with an additive $u(\cdot)$ is NP-hard \cite{Knapsack:Dantzig}.
We now show that (i) if $[p_i^\star]$ is a solution of {\perpriceprob} in (4), then a solution of GKP in \eqref{eq:GKP} is given as 
$[(\phi_i(p^\star_i)\})]$ in (1), and (ii) if $[(x_i^\star)]$ is a solution of GKP, then a solution of {\perpriceprob} is given as $p_i = c_i \cdot x_i.$ We prove (i) and (ii) by contradiction. 

\noindent{\em (i).} Let $\vec{x}^\star = \{\phi_i(p_i^\star)\}_{i \in \set{N}}$,
where $p^\star = \{p_i^\star\}_{i \in \set{N}}$ is the solution of {\perpriceprob}, i.e., utility of $p^\star$ is $u(\vec{s} \circ \vec{x}^\star)$.
Then it is straightforward to check that $\vec{x}^\star$ satisfies the constraint in \eqref{eq:GKP-condi}.
Suppose $\vec{x}^\star$ is not a solution of GKP
so that there exists a solution $\vec{x}'$ such that $u(\vec{s} \circ \vec{x}') > u(\vec{s} \circ \vec{x}^\star)$.
We now construct a {\perprice} $\pi' = \{c_i x'_i\}_{i \in \set{N}}$ from $\vec{x}'$.
 Since $\vec{x}'$ is a solution of GKP, pricing $\pi'$ satisfies the constraints in (2).
 The construction of ${\pi}'$ contradicts the optimality of $\pi^\star,$
 since the utility of $\pi'$ is strictly larger than that of $\pi^\star$.

\noindent{\em (ii).} Let $[(x_i^{\star})]$ be an optimal solution of GKP. Let $[p_i]$ be a solution of {\perpriceprob} such that $u(\vec{s} \circ [p_i]) > u(\vec{x} \circ \vec{x}^\star)$. We now construct an allocation vector $x' = [(\mathbbm{1}[p_i(s_i) \geq c_i])]$ from personalized pricing $[p_i]$. Since $[p_i]$ is budget-feasible, it is straightforward that $x'$ is a feasible solution of GKP, which again contradicts the optimality of $[(x^\star)]$, and it completes the proof. \ep

\paragraph{Proof of Theorem~\ref{thm:ex_post}} 
We first lay out the foundations in defining the ex-ante budget constraint as follows:
For $B \ge 0$ and 
$\pi \in \set{F}(B):= \{\pi : 
\sum_{i\in \set{N}} \expect{(p_i(s_i)
\mathbbm{1}[p_i(s_i) > c_i ]} \leq B
\}$,
let $\set{Q}(p, B)$ be the distribution of 
random vector $\vec{y} \in \{0,1\}^{\set{N}}$ 
consisting of $y_i = \max \big\{ \mathbbm{1}[p_i(s_i) > c_i ],
\mathbbm{1}[p_i(s_i) = c_i] b_i \big\}.$
Here, $b_i$ is an independent Bernoulli random variable with parameter $\rho (\pi, B)$, where 
\begin{align*} 
\rho(\pi,B) := \min\left\{1, 
\frac{B-\sum_{i \in {\mathcal N}} \mathbb{E} [(p_i(s_i)) \mathbbm{1}[p_i(s_i) > c_i]]}{\sum_{i \in {\mathcal N}} \mathbb{E}[(p_i(s_i)) \mathbbm{1}[p_i(s_i) = c_i]]}
\right\}  \;.
\end{align*}
Note that $\rho(p,B)\geq 0$ directly comes from the definition of $\pi$.

Thanks to the generation of $\vec{y} \sim \set{Q}(\pi,B)$
and condition on $\pi \in \set{F}(B)$,
the task allocation according to $\vec{y}$ 
verifies budget constraint in expectation sense:
\begin{align*} \textstyle
      \EXP_{\vec{y} \sim \set{Q}(p,B)} \big[
      \sum_{i \in \set{N}}
      (p_i(s_i)) y_i \big] \le B ,
\end{align*}
which is also known as 
{\em ex-ante} budget constraint.
We then formulate an ex-ante version of {\OPP}$(B)$ in \eqref{eq:opt_goal} as follows:
\begin{align}
\text{\hspace{-0.2cm} {\bf OPP-E}$(B)$:} \quad
\underset{p \in \set{F}(B)}{\textnormal{maximize}}  & \quad
V(B; p) \label{eq:ex_ante_object} 
\end{align}
where $V(B; p) := \EXP_{\vec{y} \sim \set{Q}(p, B)} [ U(\vec{s} \circ \vec{y} )]$. 

We consider a common pricing that has {\bf \em linear bonus with zero base payment,} which we call {\bf LBZB} pricing and  denote by
$\linear(\alpha)$ with the slope $\alpha \ge 0$, i.e., $p_c(\alpha) := \alpha s$. 
Next we define a constant $\alpha_B$ as the following:
\begin{align*}
&\alpha_{B} \defeq \inf \left\{ \alpha \ge 0 : 
\sum_{i \in \set{N}}\EXP 
\big[ c_i z_i (\alpha, \lambda)  \big]  \geq B,
~\exists \lambda \in [0,1]\right\},
\end{align*}
for a random variable $z_i$ defined as
$z_i(\alpha, \lambda) \defeq \mathbbm{1}[\alpha s_i > c_i]
+\lambda \mathbbm{1}[\alpha s_i = c_i].$
Now we prove that $p_c(\alpha_B)$ achieves the approximation ratio stated in the theorem.

\smallskip
We present two key lemmas whose proofs are in Section~\ref{sec:lemma}. \begin{lemma}[Ex-ante PoA]\label{lem:apx_ante} 
LBZB pricing {\linpol}$(\alpha_B)$ is $(1,\mu_B)$-approximate of $\text{OPP-E}(B)$, where $\mu_B$ is defined as follows:
\begin{align*}
& \mu_B \defeq  \frac{\sum_{i \in \set{N}}\EXP \big[  \alpha_B s_i z_i(\alpha_B, \lambda_B)
\big] }
{\sum_{i \in \set{N}}\EXP  [ c_i z_i(\alpha_B, \lambda_B)]},  \quad \text{where}\\
&\lambda_B \defeq \inf \left\{ \lambda\in [0,1] : 
\sum_{i \in \set{N}}\EXP 
\big[ c_i z_i (\alpha_B, \lambda)  \big]  \geq B \right\}.
\end{align*}
\end{lemma}  

\begin{lemma}[Connection to ex-post]\label{lem:post_to_ante}
For any $k = \omega(1)$ and $\varepsilon = o(1)$ with $k\varepsilon>2$, we have
\begin{align*}
U\!\Big(\mu_{B}B; \linear(\alpha_{B}) \Big)
\ge \Big(1-e^{-\varepsilon^2(1-\varepsilon)k/12}\Big) V\!\Big(\mu_B B ; \linear(\alpha_B) \Big) .
\end{align*}
\end{lemma}

Lemma~\ref{lem:apx_ante} provides an approximation ratio of $\linear(\alpha_B)$ to OPP-E$(B)$.
We then use Lemma~\ref{lem:post_to_ante} with $k=n^{1/2}$ and $\varepsilon = n^{-1/6}$ to translate 
the ex-ante approximation into the ex-post one, where we have
\begin{align*}
U\!\Big(\mu_{B} B; \linear(\alpha_{B}) \Big) 
&\ge \Big(1-O(e^{-n^{1/6}})\Big) V\!\Big(\mu_B B ; \linear(\alpha_B) \Big) \ge \Big(1-O(e^{-n^{1/6}})\Big) V^\star(B).
\end{align*}
This concludes the proof since $V^\star(B) \ge U\big(B;\pi^\star_p(B)\big)$ and $\mu_B = O(1)$ by assumption. \ep

\paragraph{Observation on LBZB}
We provide a couple of interpretations of LBZB introduced in the proof of Theorem~\ref{thm:ex_post}. 
We note that the linear bonus in the LBZB pricing $\linear(\alpha)$ recruits only {\em efficient} workers whose utility-to-cost ratio $s_i/c_i$ is at least $1/\alpha$, i.e., $x_i = 1$  only if $s_i / c_i \ge {1}/{\alpha}$.
A good personalized pricing may also target such efficient workers in priority, while user-agnosticism of the common pricing pays more often than the personalized one. Hence, the simple common pricing $\linear(\alpha)$ with a good choice of $\alpha = \alpha_B$ and more budget $(O(1) \times B)$ achieves at least $1 - o(1)$ of the optimal utility of {\perpriceprob}$(B)$.

We also note that one can get tighter bound on utility by optimizing the parameters in Lemma~\ref{lem:post_to_ante}, we instead use a naive parameter just to show the insight that LBZB achieves asymptotically optimal utility within constant amount of augmented budget.

For the employed LBZB common pricing to be practical, it is required that $\alpha_B$ allows easy computation, which turns to be polynomial-time computable with respect $n$ in approximate manner. To do so, 
given all the distributions of worker profiles, 
it is enough to find $\alpha$ such that $\sum_{i \in \set{N}}\EXP \big[c_i \mathbbm{1}[\alpha s_i \geq c_i] \big] \geq B$ and $\sum_{i \in \set{N}}\EXP \big[c_i \mathbbm{1}[\alpha s_i > c_i] \big] \leq B$, where of each term in the summations can be computed in $O(1)$ time. Since the available interval of $\alpha$ is bounded, using some root finding algorithms, we can find such $\alpha$ approximately in at most $c \cdot O(1) \cdot 2n$ time for some constant $c$.


\paragraph{Proof of Theorem~\ref{thm:pad_bonus}}
For a given budget $B$, we denote $u^\star(B)$ and $u_0^\star(B)$ to be the utilities of an optimal pricing by solving $\text{OPP}(B)$, and an optimal common pricing without bonus, respectively.
Given a budget $B$, we also can find $c_0$ such that $c_0n > B$.
Now, for $\varepsilon = \frac{B}{n^2c} < 1$, consider the following profiles of $n$ workers:
\begin{subnumcases}{\label{eq:eg-profile}(s_i, c_i) =}
   (s_0, c_0) & if $0 < i \leq \varepsilon n$  \nonumber\\
 (0, c_0)  &  if $\varepsilon n < i \leq n$.  \nonumber
\end{subnumcases}
For simplicity, we assume that $\varepsilon n$ is an integer without loss of generality.
Then, it is straightforward to check that $u^\star(B) = \varepsilon n s$.
Now we compute the optimal utility for pricing without personalization nor bonus under the worst-case worker arrival model. Consider a scenario where all the spammers arrive ahead of the hammers. We assume that the budget is augmented by a factor $\delta = o(n)$. Then, the total payment to spammers is computed as the following:
\begin{align*}
(1-\varepsilon)cn = nc-B/n \geq \delta B,
\end{align*}
where the inequality comes from that $\delta = o(n)$. Hence, the optimal common pricing without bonus failed to recruit any hammers, and it concludes the proof. \ep

\subsection{Proof of Lemmas}
\label{sec:lemma}

\paragraph{Proof of Lemma~\ref{lem:apx_ante}}
We prove that $V(\mu_B B ; \linear(\alpha_B))$ exactly achieves the optimum of ex-ante mechanism given budget $B$, i.e., $V(\mu_B B ; \linear(\alpha_B)) \geq V^\star(B)$.
First, since $s_i/c_i$ is also a positive random variable, 
it is straightforward to have that the right-continuity of CDF of $s_i/c_i$ implies that $\expect{c_i z_i(\alpha_B, \lambda_B)} = B$.
We now consider the following relaxed version of the ex-ante mechanism.
\begin{align*}
\text{\hspace{-0.2cm} {\bf Oracle}$(B)$:} \quad
\underset{\vec{x}= [(x_i(s_i,c_i))]\in [0,1]^n}{\textnormal{maximize}}  & \quad
\sum_{i \in \set{N}}\EXP \left[ u(s \circ x)) \right]  \\
\quad \text{\hspace{-0.2cm} s.t.}\quad & \quad
\sum_{i \in \set{N}} \EXP \left[ c_ix_i \right] \leq B.
\end{align*}
Note that $\vec{x}$ is not necessarily sampled from $\set{Q}(\vec{\pi},B)$, so that the domain of $\vec{x}$ is also extended to any real number in $[0,1]^n$. It is trivial that the optimum of \textbf{Oracle}($B$) is dominant to that of ex-ante mechanism given budget $B$.
We show that the allocation vector obtained by common pricing $\linear(\alpha_B)$ under ex-ante mechanism becomes optimal for \textbf{Oracle}($B$).
We first introduce two lemmas that are useful in the proof whose proofs are presented in the end of this section.

\begin{lemma}\label{lem:1}
Let $\vec{z}^\star$ be any optimal allocation vector for \textbf{Oracle}($B$). The following holds for any random incidence of random vector $(\vec{s}, \vec{c})$.
\begin{compactenum}[(i)]
\item If $\alpha_B s_i < c_i$, then $z_i^\star(s_i,c_i) = 0 $ a.e. for any $i \in \set{N}$.
\item If $\alpha_B s_i > c_i$, then $z_i^\star(s_i,c_i) = 1 $ a.e. for any $i \in \set{N}$.
\end{compactenum}
\end{lemma}

\begin{lemma}\label{lem:3}
For $i \in \set{N},$ let $\set{H}_i = \left\{(s_i,c_i) | \alpha_B s_i = c_i \right\}$. Let $\vec{x}$ and $\vec{x}'$ be the two random allocation vectors which satisfies 
$\sum_{i \in \set{N}} \EXP_{(s_i,c_i) \in \set{H}_i} [c_ix_i] \geq (=) \sum_{i \in \set{N}} \EXP_{(s_i,c_i) \in \set{H}_i} [c_ix_i']. $ 
Then, the expected utility of $\vec{x}$ is at least that of $\vec{x}'$ under $\set{H}_i$, i.e., 
$$
\sum_{i \in \set{N}} \EXP_{(s_i,c_i) \in \set{H}_i} [s_)x_i] \geq (=) \sum_{i \in \set{N}} \EXP_{(s_i,c_i) \in \set{H}_i} [s_ix_i']. 
$$
\end{lemma}
Let $\vec{\chi}$ be the allocation vector induced by $\linear(\alpha_B)$ under the augmented budget $\mu_B B$.
By Lemma~\ref{lem:1}, it is straightforward to check that optimal allocation vector $\vec{z}^\star$ must recruit worker $i$ if $\alpha_B s_i > c_i$, and must reject her if $\alpha_B s_i < c_i$.
We first consider the scenario where only allocating the budget to workers with $\alpha_B s_i > c_i$ fully exhaust the budget $B$.
In this case, $\sum_{i \in \set{N}} \expect{c_iz_i^{\star} \mathbbm{1}[\alpha_B s_i > c_i]}= B$, and it implies that $\vec{\chi}$ is exactly the same as $\vec{z}$ by the definition of $\alpha_B$.
Moreover, the expected payment to workers in $\linear(\alpha_B)$ can be computed as:
\begin{equation*}
\sum_{i \in \set{N}} \EXP [\alpha_B s_iz_i(\alpha_B, \lambda_B)]
= \sum_{i \in \set{N}} \bexpect{\alpha_B s_i \mathbbm{1}[\alpha_B s_i > c_i]} = \mu_B B,
\end{equation*}
which guarantees that $\linear(\alpha_B)$ is $\mu_B B$-budget feasible, and concludes the proof.
We now suppose that the budget is not exhausted, hence optimal allocation vector must recruit some workers with $\alpha_B s_i = c_i$ (workers at boundary),
i.e.,  $\sum_{i \in \set{N}} \EXP [c_iz_i^{\star} \mathbbm{1}[\alpha_B s_i > c_i]] < B$.
By the definition of $\alpha_B$ and $\lambda_B$, the following holds:
\begin{align}
\sum_{i \in \set{N}} \EXP [c_i \chi_i\mathbbm{1}[\alpha_B s_i = c_i]] &= \sum_{i \in \set{N}} \EXP[c_i \mathbbm{1}[\alpha_B s_i = c_i]]\cdot \rho(\linear(\alpha_B), \mu_B B)\\
&=\sum_{i \in \set{N}} \EXP[c_i \mathbbm{1}[\alpha_B s_i = c_i]] \cdot
 \frac{\mu_B B -\sum_{i \in \set{N}} \EXP [\alpha_B s_i \mathbbm{1}[\alpha_B s_i > c_i]]}
{ \sum_{i \in \set{N}} \EXP[\alpha_B s_i \mathbbm{1}[\alpha_B s_i = c_i]]}\\
&= \mu_B B- \sum_{i \in \set{N}} \EXP [\alpha_B s_i \mathbbm{1}[\alpha_B s_i > c_i]]. \label{lem1:cost}
\end{align}
Moreover, it is straightforward to see that any optimal allocation vector $z^\star$ must exhaust the budget in workers with $\alpha_B s_i = c_i$, i.e., 
\begin{equation}
\sum_{i \in\set{N}}\EXP [ c_i z_i^\star \mathbbm{1}[\alpha_B s_i = c_i]] 
= B - \sum_{i \in\set{N}}\EXP [ c_i z_i^\star \mathbbm{1}[\alpha_B s_i > c_i]]. \label{lem1:cost'}
\end{equation}
Now we show that the cost consumed by $\vec{\chi}$ is bigger than or equal to that by $\vec{z^\star}$. For notational simplicity, we define the random variables $\mathbbm{1}[\alpha_B s_i > c_i]$ and $\mathbbm{1} [\alpha_B s_i = c_i]$ to be $L_i$ and $E_i$, respectively, and $\rho(\linear(\alpha_B), \mu_B B)$ to be $\rho_0$.
From equation \eqref{lem1:cost},
\begin{align}
\mu_B B - \sum_{i \in \set{N}} \EXP [\alpha_B s_i L_i]&= \mu_B B - \sum_{i\in \set{N}} \EXP [\alpha_B s_i L_i] - \sum_{i \in \set{N}} \EXP [\alpha_B s_i E_i]\rho_0 + \sum_{i \in \set{N}} \EXP [\alpha_B s_i E_i]\rho_0 
\\
&= \mu_B B - \mu_B(\sum_{i \in \set{N}}\EXP [c_i L_i] + \sum \EXP [c_i E_i]\rho_0) + \sum_{i \in \set{N}} \EXP [\alpha_B s_i E_i]\rho_0 \label{lem1:cost_2}
\\
&= \mu_B (B - \sum_{i \in\set{N}}\EXP [ c_i L_i]) + \sum_{i \in \set{N}} \EXP [\alpha_B s_i E_i]\rho_0 - \mu_B \sum_{i \in \set {N}} \EXP[c_i E_i]\rho_0 \label{lem1:cost_3} 
\\
&= \mu_B (B - \sum_{i \in\set{N}}\EXP [ c_i L_i]) - \sum_{i \in \set{N}} \EXP[c_i E_i] \rho_0 (\mu_B - 1) \label{lem1:cost_4} 
\\
&\geq B - \sum_{i \in\set{N}}\EXP [ c_i L_i] \label{lem1:cost_5},
\end{align}
which is the RHS of equation~\eqref{lem1:cost'}, i.e. the expected payment in random allocation vector $\vec{z}^\star$.
Note that equation~\eqref{lem1:cost_2} holds from the definition of $\mu_B$, 
equation \eqref{lem1:cost_4} holds since $\alpha_B s_i = c_i$ under the regime that $E_i =1$,
and inequality~\eqref{lem1:cost_5} holds from the fact that $\sum_{i \in \set{N}} \EXP [c_i (L_i + \rho_0 E_i)] \leq \sum_{i \in \set{N}} \EXP [\alpha_B s_i (L_i + \rho_0 E_i)] \leq B$.
Finally by Lemma~\ref{lem:3}, we have:
\begin{eqnarray}
\sum_{i \in \set{N}} \EXP_{(s_i,c_i) \in H_i} [s_i\chi_i] \geq \sum_{i \in \set{N}} \EXP_{(s_i,c_i) \in H_i} [s_iz_i^\star], \label{thm:1}
\end{eqnarray}
which implies that the expected utility gain from workers at boundaries, i.e. workers satisfying $\alpha_B s_i = c_i$, is larger or equal on $\vec{\chi}$ than $\vec{z^\star}$.
Since it is obvious that $p_c(\alpha_B)$ is $\mu_B B$-budget feasible, combining Lemma~\ref{lem:1} and \eqref{thm:1}, 
we complete the proof of Lemma~\ref{lem:apx_ante}. \ep

\paragraph{Proof of Lemma~\ref{lem:post_to_ante}}
For simplicity, we denote LBZB pricing $p_c(\alpha_B)$ by $q$.
Let $P$ be a random variable defined by $P = \sum_{i \in \set{N}} q(s_i)$.
Now we prove that tail probability that the sum of payments to workers $P$ is larger than $(1-1/k)\mu_B B$, is lowered bounded by some exponent on $\varepsilon$ and $k$.
Since $\varepsilon = o(1)$, $\mu_B = O(1)$, and payment to any worker on LBZB is bounded above by constant asymptotically, we have $\EXP[P] \leq (1-\varepsilon)\mu_B B$ for sufficiently large $n$.
Without loss of generality, we can in fact assume that $\EXP[P] = (1-\varepsilon)\mu_B B$ by adding some dummy base payment since it can only increase the right tailed probability.
Since $k\varepsilon > 2$, we have the following inequalities:
\begin{align}
\prob{P \geq (1+\frac{\varepsilon}{2})\EXP[P]} = \prob{P \geq (1+\frac{\varepsilon}{2})(1-\varepsilon)B} &\geq \prob{P > (1-\frac{\varepsilon}{2})\mu_B B} \\
& \geq \prob{P > (1-1/k)\mu_B B}. \label{eq:lem2_cher0}
\end{align}
Now we define $Y_i = \frac{q(s_i)k}{\mu_B B}$ and $Y = \sum_{i \in \set{N}}Y_i$. 
Since $k = \omega(1)$, one can easily get $Y_i \in [0,1]$ for sufficiently large $n$, where we have the following inequality:
\begin{align}
\prob{P \geq (1+\frac{\varepsilon}{2})\EXP[P]}
&= \prob{P\frac{k}{\mu_B B}\geq \frac{k}{\mu_B B}(1+\frac{\varepsilon}{2})\EXP[P]}
= \prob{Y \geq (1+\frac{\varepsilon}{2})\EXP[Y]}
\\
&\leq \exp(-\frac{\varepsilon^2/4}{2+\varepsilon/2}\cdot \EXP[Y]) 
= \exp(-\frac{\varepsilon^2/4}{2+\varepsilon/2}\cdot \frac{k}{\mu_B B}\EXP[P])\label{eq:lem2_cher1}
\\
&= \exp(-\frac{\varepsilon^2/4}{2+\varepsilon/2}\cdot \frac{k}{\mu_B B}(1-\varepsilon)\mu_B B) \leq \exp(-\frac{\varepsilon^2(1-\varepsilon) k}{12})\label{eq:lem2_cher2}.
\end{align}
Inequality in \eqref{eq:lem2_cher1} follows from Chernoff bound and \eqref{eq:lem2_cher2} follows from $\varepsilon \leq 1$.
Hence by \eqref{eq:lem2_cher0} and \eqref{eq:lem2_cher2}, we have the following inequality:
\begin{align}
\prob{P \leq (1-1/k)\mu_B B} \geq 1 - \exp(-\varepsilon^2(1-\varepsilon) k/12).\label{eq:lem2_cher3}
\end{align}
Next, in ex-post mechanism with augmented budget $\mu_B B$, we claim that the probability that any worker is offered the pricing policy (not discarded by the ex-post budget constraint) is at least $\prob{P \leq (1-1/k)\mu_B B}$, i.e. the probability that the sum of the payments to all the workers is at most $(1-1/k)\mu_B B$.
This is quite obvious since if the sum of payments to all the workers is at most $(1-1/k)\mu_B B $ for any random incidence, then the budget remains in this moment and any worker will be offered the pricing policy.
Hence, we have the following inequalities on the expected utility of LBZB in ex-post mechanism:
\begin{align}
U(\mu_B B;\linear(\alpha_B))&=\EXP_{\vec{x}\sim\set{P}(\linear(\alpha_B),B,\sigma)}[u(\vec{s} \circ \vec{x} )]
\\
&\geq \sum_{i \in \set{N}}\bexpect{s_i \cdot z_i(\alpha_B, \lambda_B)}\prob{ P \leq (1-1/k) B}
\\
&\geq \sum_{i \in \set{N}}\bexpect{s_i \cdot z_i(\alpha_B, \lambda_B) }
\lf(1 - \exp(-\varepsilon^2(1-\varepsilon) k/12)\ri)
\\ 
&= \lf(1 - \exp(-\varepsilon^2(1-\varepsilon) k/12)\ri)V(\mu_B B; \linear(\alpha_B)),
\end{align}
and it concludes the proof.\ep

\paragraph{Proof of Lemma~\ref{lem:1}}
The underlying intuition is that greedy algorithm outputs the exact optimum of the {\em fractional} knapsack problem.
We use proof by contradiction. 
Let the sample space of the random vector $(s_i, c_i)$ be $O_i$. 
We define the following sets.
\begin{align*}
\set{E}_i &:= \{ (\vec{s}, \vec{c}) \; | \; \alpha_B s_i < c_i, z_i^\star(\vec{s}, \vec{c}) > 0 \}\\
\set{F}_i &:= \{ (\vec{s}, \vec{c}) \; | \; \alpha_B s_i \geq c_i, z_i^\star(\vec{s}, \vec{c}) < 1 \}\\
\set{G}_i &:= \{ (\vec{s}, \vec{c}) \; | \; \alpha_B s_i \geq c_i, z_i^\star(\vec{s}, \vec{c}) = 1 \}.
\end{align*}
Note that given any optimal allocation vector $z^\star(\vec{s}, \vec{c})$, $E_i$ denotes the set of all incidences of random vector $(\vec{s}, \vec{c})$ that satisfies $\alpha_B s_i < c_i$ and $z^\star_i(\vec{s}, \vec{c}) > 0$.
Firstly, we claim that if there exists $i \in \set{N}$ such that $\bprob{\set{E}_i} > 0,$ then 
there exist $j \in \set{N}$ such that $\bprob{\set{F}_j} > 0$.
Suppose not, i.e. $\bprob{\set{E}_i} > 0$ for $i \in \set{N}$, and $\bprob{\set{F}_j} = 0$ for any $j \in \set{N}$.
By assumption, we get $\bprob{\set{G}_j} = \bprob{\alpha_B s_i \geq c_i}$.
Since $\set{E}_i$, $\set{F}_i$, and $\set{G}_i$ are mutually exclusive, the expected payment to workers in $\vec{z}^\star$ satisfies the following inequalities:
\begin{align}
B&\geq \sum_{i \in \set{N}} \EXP_{(s_i,c_i) \in \set{E}_i} [c_i z_i^\star] + \EXP_{(s_i,c_i) \in \set{F}_i} [c_i z_i^\star] + \EXP_{(s_i,c_i) \in \set{G}_i} [c_i z_i^\star] \\
&\geq \sum_{i \in \set{N}} \EXP_{(s_i,c_i) \in \set{E}_i} [c_i z_i^\star]  + \EXP_{(s_i,c_i) \in \set{G}_i} [c_i z_i^\star] \\
&\geq \sum_{i \in \set{N}}\EXP_{(s_i,c_i) \in \set{E}_i} [c_i z_i^\star]  + \EXP [c_i \mathbbm{1}[\alpha_B s_i \geq c_i]]  \\
&\geq \sum_{i \in \set{N}}\EXP_{(s_i,c_i) \in \set{E}_i} [c_i z_i^\star] + B > B,
\end{align}
which is a contradiction.
Hence we conclude that if there exists $i \in \set{N}$ with $\bprob{\set{E}_i} > 0,$ then there exists $j \in \set{N}$ with $\bprob{\set{F}_j} > 0$.
By our claim, we can choose two random incidences $(\bar{s_i}, \bar{c_i}) \in \set{E}_i$ and $(\tilde{s_j},\tilde{c_j}) \in \set{F}_j$, which are $i$-th and $j$-th element in incidences of random vector $(\bar{\vec{s}}, \bar{\vec{c}})$ and $(\tilde{\vec{s}}, \tilde{\vec{c}})$, respectively.
Now we construct an allocation vector $\vec{y} \in [0,1]^n$, which is equal to $\vec{z}^\star$ except the two sample points $(\bar{\vec{s}}, \bar{\vec{c}})$ and $(\tilde{\vec{s}}, \tilde{\vec{c}})$ so that $y_j(\tilde{s_j},\tilde{c_j}) = z_j^{\star}(\tilde{s_j}, \tilde{c_j}) + \varepsilon$, and 
$y_i(\bar{s_i}, \bar{c_i}) = z_i^{\star}(\bar{s_i}, \bar{c_i}) - \frac{\tilde{c_j}}{\bar{c_i}}\varepsilon$, where $\varepsilon = \min \left\{ \frac{\bar{c_i}}{\tilde{c_j}}z_i^{\star}(\bar{s_i}, \bar{c_i}) , 1- z_j^{\star}(\tilde{s_j}, \tilde{c_j})  \right\}$.
Then, the difference between the total expected payment to workers in $\vec{y}$ and $\vec{z}^\star$ can be computed as the following:
\begin{align*}
\sum_{i \in \set{N}}\EXP[c_i z_i^\star] - \sum_{i \in \set{N}} \EXP[c_i y_i] &= \left (\tilde{c_j}y_j(\tilde{s_j}, \tilde{c_j}) + \bar{c_i}y_i(\bar{s_i}, \bar{c_i}) \right) -
\left ( (\tilde{c_j}z_j^\star(\tilde{s_j}, \tilde{c_j}) +\bar{c_i}z_i^\star(\bar{s_i}, \bar{c_i}))\right)
\\
&=
\tilde{c_j}(z_j^\star(\tilde{s_j}, \tilde{c_j}) + \varepsilon) + \bar{c_i}(z_i^\star(\bar{s_i}, \bar{c_i})- \frac{\tilde{c_j}}{\bar{c_i}}\varepsilon)-(\tilde{c_j}z_j^\star(\tilde{s_j}, \tilde{c_j}) +\bar{c_i}z_i^\star(\bar{s_i}, \bar{c_i}))
\\
&=
\tilde{c_j}\varepsilon - \bar{c_i}\cdot \frac{\tilde{c_j}}{\bar{c_i}} \varepsilon = 0,
\end{align*}
which implies that $\vec{y}$ is also a feasible point for \textbf{Oracle}($B$). 
Now we compute the difference of expected utility between $\vec{y}$ and $\vec{z}^\star$ as the following:
\begin{align*}
\sum_{i \in \set{N}}\EXP[s_iy_i] - \sum_{i \in \set{N}} \EXP[s_i z_i^\star]&=
\tilde{s_j }(z_j^\star(\tilde{s_j}, \tilde{c_j}) + \varepsilon) + \bar{s_i }(z_i^\star(\bar{s_i}, \bar{c_i}) -  \frac{\tilde{c_j}\varepsilon}{\bar{c_i}}) - \Big(\tilde{s_j }z_j^\star(\tilde{s_j}, \tilde{c_j}) +
\bar{s_i }z_i^\star(\bar{s_i}, \bar{c_i})\Big) \\
&=\tilde{s_j }\varepsilon -\bar{s_i }\frac{\tilde{c_j}}{\bar{c_i}}\varepsilon > \frac{\varepsilon}{\alpha_B}(\tilde{c_j} - \bar{c_i}\frac{\bar{c_i}}{\bar{c_i}}) = 0,
\end{align*}
which implies that the expected utility of $\vec{y}$ is strictly larger than $\vec{z}.$ This contradicts the assumption that $\vec{z}$ is optimal solution for \textbf{Oracle}($B$), and it completes the proof of Lemma 4.3. Since the proof for statement (ii) is exactly analogous to that of (i), we omit the detailed proof.\ep

\paragraph{Proof of Lemma~\ref{lem:3}} 
The following inequality directly completes the proof.
\begin{align}
\hspace{-0.1cm}\sum_{i \in \set{N}} \EXP_{(s_i,c_i) \in \set{H}_i} [s_ix_i] = \sum_{i \in \set{N}} \expect{s_ix_i \mathbbm{1}[ \alpha_B s_i = c_i]}  & =\sum_{i \in \set{N}} \bexpect{\frac{c_i}{\alpha_B}x_i \mathbbm{1}[ \alpha_B s_i = c_i]} \label{eq:lem3_1} \\
&\geq \sum_{i \in \set{N}} \bexpect{\frac{c_i}{\alpha_B}x_i' \mathbbm{1}[ \alpha_B s_i = c_i]}    \label{eq:lem3_2} \\
&= \sum_{i \in \set{N}} \EXP_{(s_i,c_i) \in \set{H}_i} [s_ix_i'], \label{eq:lem3_3}  
\end{align}
where \eqref{eq:lem3_1} holds from the definition of $\set{H}_i,$ 
and \eqref{eq:lem3_2} and \eqref{eq:lem3_3} from our assumption.
It is also straightforward to check the equality version of the lemma from \eqref{eq:lem3_2}.  \ep

\subsection{Proof of Propositions}
\label{sec:propositions}
\paragraph{Proof of Proposition~\ref{prop:1}}
In ratio-equivalent regime, it is straight forward to check that $\mu_B$ introduced in the proof of Lemma~\ref{lem:apx_ante} is equal to $1$ for $p_{lin}(k)$.
Hence from Theorem~\ref{thm:ex_post}, we conclude that $p_{lin}(k)$ is $(1,1)-$approximate as $n$ goes to infinity.\ep

\paragraph{Proof of Proposition~\ref{prop:2}}
In the proof of Theorem~\ref{thm:ex_post}, we have that $p_c(\alpha_B)$ achieves $(1,\mu_B)-$approximate in ex-ante version. We now claim that $p_{thre}(c, \beta_B)$ achieves the same utility approximation ratio without any augmented budget in ex-ante version.
By our assumption on cost equivalence, the following holds for the definition of $\alpha_B$:
\begin{align*}
\alpha_{B} &\defeq \inf \left\{ \alpha \ge 0 : 
\sum_{i \in \set{N}}\EXP 
\big[ c_i z_i (\alpha, \lambda)  \big]  \geq B,
~\exists \lambda \in [0,1]\right\} \\
&= \inf \left\{ \alpha \ge 0 : 
c \sum_{i \in \set{N}}\EXP 
\big[  z_i (\alpha, \lambda)  \big]  \geq B,
~\exists \lambda \in [0,1]\right\} \\
&= \inf \left\{ \alpha \ge 0 : 
c \sum_{i \in \set{N}}\prob{\alpha s_i \geq c} \geq B\right\}
=c\cdot \beta_B
\end{align*}
Hence from Lemma~\ref{lem:1}, we find that $p_{thre}(c, \beta_B)$ under ex-ante constraint recruits every workers with $\alpha_B s_i \geq c_i$ as the same as optimal offline algorithm does.
Next, the expected payment for the workers at boundary, i.e. workers with $\alpha_B s_i = c_i$, in $p_{thre}(c, \beta_B)$ under ex-ante constraint is the same as that of optimal offline algorithm as follow:
\begin{align*}
\rho(p_{thre}(c, \beta_B), B) \cdot c\sum_{i \in\set{N}}\EXP [\mathbbm{1}[\alpha_B s_i = c]] 
&= B - c\sum_{i \in\set{N}}\EXP [\mathbbm{1}[\alpha_B s_i > c]]\\
&= B - \sum_{i \in\set{N}}\EXP [c_i z_i^\star \mathbbm{1}[\alpha_B s_i > c_i]]\\
&= \sum_{i \in\set{N}}\EXP [c_i z_i^\star \mathbbm{1}[\alpha_B s_i = c_i]].
\end{align*}
Hence by Lemma~\ref{lem:3}, we find that the expected utility from both $p_{thre}(c, \beta_B)$ and optimal offline algorithm for the workers at boundary are exactly the same.
Finally, $p_{thre}(c, \beta_B)$ achieves the exact optimum in Oracle problem, and combined with the analogous statement in Lemma~\ref{lem:post_to_ante}, $p_{thre}(c, \beta_B)$ is $(1,1)-$approximate in asymptotic manner.\ep

\paragraph{Proof of Proposition~\ref{prop:3}}
Let $p(s_i)$ be any common pricing equipped with non-zero bonus function.
Since the support of $s_i$ is a singleton, the support of $p(s_i)$ for $i \in \set{N}$ is also a singleton, and we let that element be $\nu_i$. Then, it is straightforward to check that $p'(s_i) = \nu$ achieves the exact same expected utility as $p(s_i)$, where it concludes the proof.\ep